\def\year{2020}\relax
\documentclass[letterpaper]{article} 
\usepackage{aaai20}  
\usepackage{times}  
\usepackage{helvet} 
\usepackage{courier}  
\usepackage[hyphens]{url}  
\usepackage{graphicx} 
\usepackage{amsmath}
\usepackage{amssymb}
\usepackage{subfigure}
\usepackage{graphicx}
\usepackage[dvipsnames]{xcolor}
\usepackage{algorithm}  
\usepackage{algpseudocode}  
\usepackage{booktabs}
\usepackage{multirow}

\makeatletter
\newenvironment{breakablealgorithm}
  {
   \begin{center}
     \refstepcounter{algorithm}
     \hrule height.8pt depth0pt \kern2pt
     \renewcommand{\caption}[2][\relax]{
       {\raggedright\textbf{\ALG@name~\thealgorithm} ##2\par}%
       \ifx\relax##1\relax 
         \addcontentsline{loa}{algorithm}{\protect\numberline{\thealgorithm}##2}%
       \else 
         \addcontentsline{loa}{algorithm}{\protect\numberline{\thealgorithm}##1}%
       \fi
       \kern2pt\hrule\kern2pt
     }
  }{
     \kern2pt\hrule\relax
   \end{center}
  }
\makeatother

\title{Bayesian Symbolic Regression}
\author{Ying Jin\textsuperscript{\rm 1}, Weilin Fu\textsuperscript{\rm 1}, Jian Kang\textsuperscript{\rm 2}, Jiadong Guo\textsuperscript{\rm 1}, Jian Guo\textsuperscript{\rm 1}\\
\textsuperscript{\rm 1}Peng Cheng Laboratory\\
\textsuperscript{\rm 2}{University of Michigan, Department of Biostatistics}}
\begin{document}

\maketitle

\begin{abstract}
Interpretability is crucial for machine learning in many scenarios such as quantitative finance, banking, healthcare, etc. Symbolic regression (SR) is a classic interpretable machine learning method by bridging X and Y using mathematical expressions composed of some basic functions. However, the search space of all possible expressions grows exponentially with the length of the expression, making it infeasible for enumeration. Genetic programming (GP) has been traditionally and commonly used in SR to search for the optimal solution, but it suffers from several limitations, e.g. the difficulty in incorporating prior knowledge in GP; overly-complicated output expression and reduced interpretability etc.

To address these issues, we propose a new method to fit SR under a Bayesian framework. Firstly, Bayesian model can naturally incorporate prior knowledge (e.g., preference of basis functions, operators and raw features) to improve the efficiency of fitting SR. Secondly, to improve interpretability of expressions in SR, we aim to capture concise but informative signals. To this end, we assume the expected signal has an additive structure, i.e., a linear combination of several concise expressions, of which complexity is controlled by a well-designed prior distribution. In our setup, each expression is characterized by a symbolic tree, and therefore the proposed SR model could be solved by sampling symbolic trees from the posterior distribution using an efficient Markov chain Monte Carlo (MCMC) algorithm. Finally, compared with GP, the proposed BSR(\textbf{B}ayesian \textbf{S}ymbolic \textbf{R}egression) method doesn't need to keep an updated ``genome pool'' and so it saves computer memory dramatically.

Numerical experiments show that, compared with GP, the solutions of BSR are closer to the ground truth and the expressions are more concise. Meanwhile we find the solution of BSR is robust to hyper-parameter specifications such as the number of trees in the model. 

\end{abstract}

\section{Introduction}

Symbolic regression is a special regression model which assembles different mathematical expressions to discover the association between the response variable and the predictors, with applications studied in \cite{681044}, \cite{10.2166/hydro.1999.0010}, \cite{DAVIDSON200395}, etc. Without a pre-specified model structure, it is challenging to fit symbolic regression, which requires to search for the optimal solution in a large space of mathematical expressions and estimate the corresponding parameters simultaneously. 

Traditionally, symbolic regression is solved by combinatorial optimization methods like Genetic Programming  (GP) that evolves over generations, see \cite{4632147}, \cite{dabhi2011empirical}, \cite{chen2017feature}, \cite{RePEc:tiu:tiutis:65a72d10-6b09-443f-8cb9-88f3bb3bc31b}, etc. However, GP suffers from high computational complexity and overly complicated output expressions, and the solution is sensitive to the initial value, see \cite{Korns2011}. Some modifications of the original GP algorithm have been proposed to address those problems including~\cite{AmirHaeri:2017:SGP:3162421.3162533}~which incorporates statistical information of generations, \cite{McConaghy2011} which deterministically builds higher-level expressions from 'elite' building blocks, \cite{6557774} which employs a hybrid of GP and deterministic methods, \cite{article} uses a divide and conquer strategy to decompose the search space and reduce the model complexity, and \cite{localOp} which proposes a local optimization method to control the complexity of symbolic regression.

Although some efforts have been made to improve GP, its intrinsic disadvantages still remain unsolved. Some research work explores SR estimation methods other than GP. For example, \cite{OLIVETTIDEFRANCA201818} which introduces a new data structure called Interaction-Transformation to constrain the search space and simplify the output symbolic expression, \cite{McConaghy2011} which uses pathwise regularized learning to rapidly prune a huge set of candidate basis functions down to compact models, \cite{DBLP:journals/corr/ChenLJ17} assumes regression models are spanned by a number of elite bases selected and updated by their proposed algorithm, \cite{DBLP:journals/corr/abs-1904-03368} introduces a neuro-encoded expression programming with recurrent neural networks to improve smoothness and stability of the search space, \cite{DBLP:journals/corr/abs-1901-07714}which introduces an expression generating neural network and proposes an Monte Carlo tree search algorithm to produce expressions that match given leading powers.

In this work, we consider to fit symbolic regression under a Bayesian framework, which can naturally incorporate prior knowledge, can improve model interpretability and can potentially simplify the structure and find prominent components of complicated signals. The key idea is to represent each mathematical expression as a symbolic tree, where each child node denotes one input value and the parent node denotes the output value of applying the mathematical operator to all the input values from its child nodes. To control model complexity, the response variable $y$ is assumed to be a linear combination of multiple parent nodes whose descendant nodes (or leaf nodes) are the predictor $\mathtt{x}$. We develop a prior model for the tree structures and assign informative priors to the associated parameters. Markov chain Monte Carlo (MCMC) methods are employed to simulate the posterior distributions of the underlying tree structures which correspond to a combination of multiple mathematical expressions. 

The paper is organized as follows. First, we present our Bayesian symbolic regression model by introducing the tree representation of mathematical expressions. Then we develop an MCMC-based posterior computation algorithm for the proposed model. Finally, we demonstrate the superiority of the proposed method compared to existing alternatives via numerical experiments. 

In the following parts, we will refer to our symbolic regression method based on Bayesian framework as Bayesian Symbolic Regression or BSR in exchange.

\section{Bayesian Symbolic Regression with Linearly-Mixed Tree Representations}

Denote by $\mathbf{x}=(x_1,\dots,x_d)\in \mathbb{R}^d$ the predictor variables and by $y\in \mathbb{R}$ the response variable. We consider a symbolic regression model:
$$y = g(\mathbf{x}) + \epsilon,$$
where $g(\cdot)$ is a function represented by a combination of mathematical expressions taking predictors $\mathbf{x}$ as the input variable. Specifically, the mathematical operators such as $+$, $\times$, $\ldots$, and arithmetic functions like $\exp(\cdot), \cos(\cdot),\ldots $,  can be in the search space of mathematical expressions. For example, $g(\mathbf{x})=x_1+2\cos(x_2)+\exp(x_3)+0.1$.

\subsection{Choice of Basic Operators}
All possible mathematical expressions are combinations of elements in a set of basic functions. The choice of basic operators is a building block of our tree representation, see \cite{Nicolau2018OnTE}. In this paper, we adopt the commonly-used operators $+$, $\times$, \texttt{exp()}, \texttt{inv}$(x)=1/x$, \texttt{neg}$(x)=-x$ and linear transformation \texttt{lt}$(x)=ax+b$ with parameters $(a,b)\in\mathbb{R}^2$. They are able to express $-$ and $\div$ with symmetric binary operators. In practice, the basic operators can be specified by users. 

\subsection{From Expressions to Trees}

The mathematical expression can be equivalently represented by a tree denoted by $T$, with non-terminal nodes indicating operations and terminal nodes indicating the selected features. $T$ is a binary tree but not necessarily a complete tree.

Specifically, a non-terminal node has one child node if it is assigned a unary operator, and two if assigned a binary operator. For example, a non-terminal node with operator $+$ represents the operation that the values of its two child nodes are added up. For a non-terminal unary operator, for example \texttt{exp()}, it means taking exponential of the value of its child node. Note that some operators may also be associated with parameters, like linear transformation $\texttt{lt}(x)=ax+b$ with parameters $(a,b)\in\mathbb{R}^2$. We collect these parameters in a vector $\Theta$.

\begin{figure}
\centering
\includegraphics[height=1in]{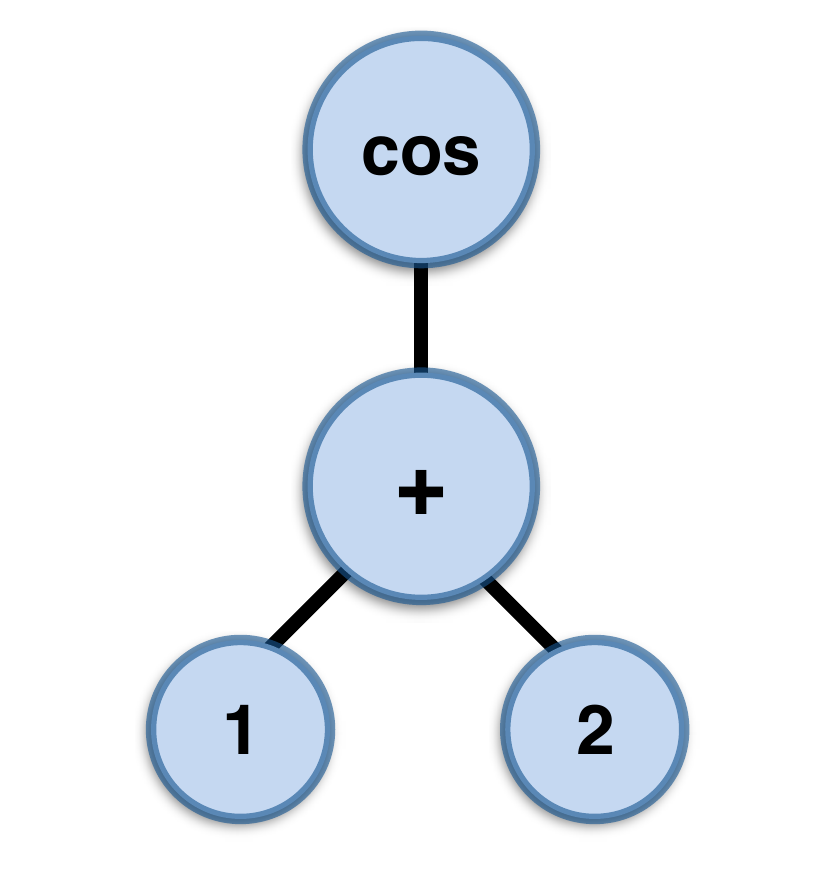}
\caption{Tree representation of $\cos(x_1+x_2)$}
\label{fig:tree}
\end{figure}

On the other hand, each terminal node $\eta$ specified by $i_k\in M$ represents a particular feature $x_{i_k}$ of the data vector. Here $M$ is the vector including features of all terminal nodes. For a tree of depth $d$, we start from the terminal nodes by performing the operations indicated by their parents, then go to their parents and perform upper-level operations accordingly. We obtain the output at the root node. For example, the tree in Figure \ref{fig:tree} represents $g(\mathbf{x})=\cos(x_1+x_2)$, which consists of two terminal nodes $1,2$ and two non-terminal nodes $\cos$, $+$.

In short, the tree structure $T$ is the set of nodes $T=(\eta_1,\dots,\eta_t)$, corresponding to operators with zero to two child nodes. Some operators involve parameters aggregated in $\Theta$. From predictor $\mathbf{x}$, terminal nodes select features specified by $M=(i_1,\dots,i_p)$, where $i_k$ indicates adopting $x_{i_k}$ of vector $\mathbf x$ as the input of the corresponding node $\eta_k$. The specification of $T$, $\Theta$ and $M$ represents an equivalent tree for a mathematical expression $g(\cdot;T,M,\Theta)$.

\subsection{Priors on Tree Representations}
Under a Bayesian modeling framework, it is critical to specify appropriate priors for parameters, as it has the flexibility to incorporate prior knowledge to facilitate more accurate posterior inferences.  In our model, we are interested in making inferences on the tree structure $T$, the parameter $\Theta$ and the selected feature indices $M$.  

To ensure the model interpretability, we aim to control the size of tree representations, or equivalently, the complexity of mathematical expressions. The default prior of operators and features are uniform distributions, indicating no preference for any particular operator or feature. They can be user-specified weight vectors to pose preferences.

For a single tree, we adopt prior distributions on $T$, $M$ and $\Theta$ in a similar fashion as those for Bayesian regression tree models in \cite{10.2307/2669832} as follows. Of note, although the prior models are similar, our model and tree interpretations are completely different from the Bayesian regression tree model.  

\subsubsection{Prior of Tree Structure $T$}
We specify the prior $p(T)$ by assigning the probabilities to each event in the process of constructing a specific tree. The prior construction starts from the root node.

A node is randomly assigned a particular operator according to the prior. The operator indicates whether it extends to one child node, or split into two child nodes, or function as a terminal node. Starting from the root, such growth performs recursively on newly-generated nodes until all nodes are assigned operators or terminated.

Specifically, for a node with depth $d_\eta$, i.e. the number of nodes passed from it to the node, with probability
$
p_{1}(\eta,T) =  \alpha (1+d_\eta)^{-\beta}
$.
It is a non-terminal node, which means it has descendants. Here $\alpha$, $\beta$ are prefixed parameters that guides the general sizes of trees in practice. The prior also includes a user-specified basic operator set and a corresponding weight vector indicating the probabilities of adopting each operator for a newly-grown node. For example, we specify the operator set (operator) as Ops$=($\texttt{exp()}, \texttt{lt()}, \texttt{inv()}, \texttt{neg()}, +, $\times)$ where \texttt{lt}$(x)=ax+b$, \texttt{inv}$(x)=1/x$, \texttt{neg}$(x)=-x$, and the uniform weight vector $w_{op}=(1/6,1/6,1/6,1/6,1/6,1/6)$. Such default choice shows no preference for any particular operator.

With probability $p_1(\eta,T)$, the node $\eta$ is assigned an operator according to $w_{op}$ if it is non-terminal and grows its one or two child nodes. Then its child nodes grow recursively. Otherwise it is a terminal node and assigned some feature in a way specified later. The construction of a tree is completed if all nodes are assigned or terminated.

\subsubsection{Prior of Terminal Nodes $M$}
When a node is terminated, it is assigned a feature of $\mathbf{x}$ according to the prior of features as input of the expression. The number and locations of terminal nodes are decided by structure of $T$. Conditioned on $T$, the specific feature that one terminal node takes is randomly generated with probabilities indicated by weight vector $w_{ft}$. The default choice is uniform among all features, i.e., $w_{ft}=(1/d,\dots,1/d)$. It can also be user-specified to highlight some preferred features.

\subsubsection{Prior of \texttt{lt()} Parameters}
An important operator we adopt here is linear transformation \texttt{lt}$(x)=ax+b$ associated with linear parameters $(a,b)\in \mathbb{R}^2$. \texttt{lt()} includes scalings and well enriches the set of potential expressions. Such operation is discussed in \cite{10.1007/3-540-36599-0_7} and proved to improve the fitting. Pairs of linear parameters $(a,b)$ are assembled in $\Theta$ and are considered independent. 

Let $L(T)$ be the set of \texttt{lt()} nodes in $T$, and each node $\eta$ is associated with parameters $(a_\eta,b_\eta)$, then the prior of $\Theta$ is
\begin{small}
$$
p(\Theta \mid T) = \prod_{\eta\in L(T)} p(a_\eta,b_\eta),
$$
\end{small}
where $a_\eta$'s, $b_\eta$'s are independent and
\begin{small}
$$
a_\eta \sim N(1,\sigma_a^2),\quad b_\eta \sim N(0,\sigma_b^2).
$$
\end{small}
This indicates that the prior of the linear transformation is a Gaussian and centered around identity function. The prior of $\sigma_\Theta = (\sigma_a,\sigma_b)$ is conjugate prior of normal distribution, which is
\begin{small}
$$
\sigma_a^2 \sim IG(\nu_a/2,\nu_a \lambda_a /2),\quad \sigma_b^2 \sim IG(\nu_b/2,\nu_b \lambda_b /2),
$$
\end{small}
where $\nu_a$, $\lambda_a$, $\nu_b$, $\lambda_b$ are pre-specified hyper-parameters.

\subsection{Find the Signal: Linear Mixture of Simpler Trees}

Many popular machine learning techniques, such as neural networks, can approximate functions very well, but  they are difficult to interpret.  A widely celebrated advantage of symbolic regression is its interpretability and good performance of approximating functions.  The model fitting of symbolic regression usually results in relatively simple mathematical expressions, it is straightforward to understand the relationship between the predictors $\mathbf{x}$ and the response variable $y$. 

However, if symbolic regression produces too complicated expressions, the interpretation of the model fitting becomes challenging: there exists a tradeoff between simplicity and accuracy. To highlight the superiority of symbolic regression in interpretability over other methods, we aim at finding the most prominent and concise signals. If the features are strong and expressive, we assume that the expression should not involve too many features, and the transformation should not be too complicated.

Moreover, the real-world signal may be a combination of simple signals, where only a small amount of simpler ones play a significant role. A simpler idea has its roots in \cite{Keijzer2004}, where the output is appropriately scaled. SR has also been addressed with methods related to generalized linear models, summarized in \cite{DBLP:journals/corr/ZegklitzP17}.

In this sense, we model the final output $y$ to be centered at some linear combination of relatively simple expressions
\begin{small}
\begin{equation*}
y = \beta_0 + \sum_{i=1}^{k} \beta_i \cdot g(\mathbf{x};T_i,M_i,\Theta_i) + \epsilon, \quad \epsilon \sim N(0,\sigma^2)
\end{equation*}
\end{small}
where $k$ is a pre-specified number of simple components,  $g(\mathbf{x};T_i,M_i,\Theta_i)$ is a relatively simple expression represented by a symbolic tree, and $\beta_i$ is the linear coefficient for the $i$-th expression. The coefficients $\beta_i$, $i=0,\dots,k$ is obtained by OLS linear regression using intercept and $g(\cdot;T_i,M_i,\Theta_i), i=1,\dots,k$. Let $\{(T_i,M_i,\Theta_i)\}_{i=1}^k$ denote the series of tuples $(T_i,M_i,\Theta_i)$, $i=1,\dots,k$. Let $OLS()$ denote the OLS fitting result, then a simpler form is 
\begin{small}
\begin{equation*}
y = OLS\big(x,\{(T_i,M_i,\Theta_i)\}_{i=1}^k\big) + \epsilon, \quad \epsilon \sim N(0,\sigma^2)
\end{equation*}
\end{small}
where the prior of the noise scale is the conjugate inverse gamma distribution
\begin{small} 
\begin{equation*}
\sigma^2 \sim IG(\nu/2,\nu\lambda/2)
\end{equation*}
\end{small}
where $\nu$ and $\lambda$ are pre-specified parameters. Additionally let $(T,M,\Theta)=\{(T_i,M_i,\Theta_i)\}_{i=1}^k$, the joint likelihood is 
\begin{small}
\begin{align*}
&p(y,(T,M,\Theta),\sigma,\sigma_\Theta \mid x) \\
=& p(y\mid OLS\big(x,T,M,\Theta\big),\sigma^2) p(M,T) p(\Theta\mid T,
\sigma_\Theta^2) p(\sigma_\Theta^2) p(\sigma^2) \\
=& p(y\mid OLS\big(x,T,M,\Theta\big),\sigma^2)p(\sigma^2) \times \prod_{i=1}^{k} p(M_i\mid T_i) p(T_i)  p(\Theta_i \mid T_i,\sigma_\Theta^2).
\end{align*}
\end{small}

\section{Posterior Inference}

We employ the Metropolis-Hastings (MH) algorithm proposed in \cite{metropolis1953equation} and \cite{10.2307/2334940} to make posterior inferences on the proposed model. Note that $(T,M,\Theta)$ represents the set of $k$ trees $\{T_i,M_i,\Theta_i\}_{i=1}^k$, and $(T^s,M^s,\Theta^s)$ denotes the set of $k$ trees that the MH algorithm accepts at the $s$-th iteration.

With a pre-specified number of trees $k$, our method modifies the structure of the $i$-th tree by sampling from the proposal $q(\cdot\mid \cdot)$, and accepts the new structure with probability $\alpha$, which can be calculated according to MH algorithm. Otherwise the $i$-th tree stays at its original form. The $k$ trees are updated sequentially, so to illustrate, we first show how a single tree is modified at each time.

The sampling of a new tree consists of three parts. The first is the structure specified by $T$ and $M$, which is discrete. Here $T$ and $M$ stand for a single tree. The second part is $\Theta$ aggregating parameters of all \texttt{lt()} nodes. The dimensionality of $\Theta$ may change with $(T,M)$ since the number of \texttt{lt()} nodes vary among different trees. To address the trans-dimensional problem, we use the reversible jump MCMC algorithm proposed by \cite{green1995reversible}. For simplicity, denote by $S=(T,M)$ the structure parameters. The third part is sampling $\sigma^{2}$ from an inverse gamma prior.

\subsection{Structure Transition Kernel}
We first specify how the sampling algorithm jumps from a tree structure to a new one. Inspired by \cite{10.2307/2669832} and considering the nature of calculation trees, we design the following seven reversible actions. The probabilities from $S=(T,M)$ to new structure $S^*=(T^*,M^*)$ is denoted as the proposal $q(S^*\mid S)$.

\begin{itemize}
    \item \textbf{Stay:} If the expression involves $n_l\geq 0$ \texttt{lt()} operators, with probability $p_0 = n_l/4(n_l+3)$, the structure $S=(T,M)$ stays unchanged, and ordinary MH step follows to sample new linear parameters.
    
    \item \textbf{Grow:} Uniformly pick a terminal node and activate it. A sub-tree is then generated iteratively, where each time a node is randomly terminated or assigned an operator from the prior until all nodes are terminated or assigned.
    
    To regularize the complexity of the expression, the proposal grows with lower probability when the tree depth and amount of nodes are large. The probability of Grow is
    $
    p_g = \frac{1-p_0}{3}\cdot \min \big\{1, \frac{8}{N_{nt}+2}\big\}
    $
    ,where $N_{nt}$ is the number of non-terminal nodes.
    
    \item \textbf{Prune:} Uniformly pick a non-terminal node and turn it into a terminal node by discarding its descendants. Then randomly choose a feature of $\mathbf{x}$ to the newly pruned node. 
    
    We set the probability of Prune as
    $
    p_p = \frac{1-p_0}{3} - p_g
    $
    such that Grow and Prune share one-third of the probability that the structure does not Stay.
    
    \item \textbf{Delete:} Uniformly pick a candidate node and delete it. 

    Specifically, the candidate should be non-terminal. Also, if it is a root node, it needs to have at least one non-terminal child node to avoid leaving a terminal node as the root node. If the picked candidate is unary, then we just let its child replace it. If it is binary but not root, we uniformly select one of its children to replace it. If the picked candidate is binary and the root, we uniformly select one of its non-terminal children to replace it.
    
    We set the probability of Delete as
    $
    p_d =  \frac{1-p_0}{3} \cdot \frac{N_c}{N_c+3},
    $
    where $N_c$ is the number of aforementioned candidates.
    
    \item \textbf{Insert:} Uniformly pick a node and insert a node between it and its parent. The weight of nodes assigned is $w_{op}$. If the inserted node is binary, the picked node is set as left child of the new node, and the new right child is generated according to the prior.
    
    The probability of Insert is set as 
    $
    p_i =  \frac{1-p_0}{3} - p_d
    $
    such that Delete and Insert share one-third of the probability that the structure does not Stay.
    
    \item \textbf{ReassignOperator:} Uniformly pick a non-terminal node, and assign a new operator according to $w_{op}$. 

    If the node changes from unary to binary, its original child is taken as the left child, and we grow a new sub-tree as right child. If the node changes from binary to unary, we preserve the left sub-tree (this is to make the transition reversible). 

    \item \textbf{ReassignFeature:} Uniformly pick a terminal node and assign another feature with weight $w_{ft}$.
        
    The probability of ReassignOperator and ReassignFeature is set as
    $
    p_{ro} = p_{rf}= \frac{1-p_0}{6}
    $
\end{itemize}

Note that the generation of the 'tree' is top-down, creating sub-trees from nodes. However, the calculation is bottom-up, corresponding to transforming the original features and combine different sources of information. 

The above discrepancy can be alleviated by our design of proposal. \textbf{Grow} and \textbf{Prune} creates and deletes sub-trees in a top-down way, which corresponds to changing a "block", or a higher level feature represented by the sub-tree in the expression. On the other hand, \textbf{Delete} and \textbf{Insert} modify the higher-level structure by changing the way such "blocks" combine and interact in a bottom-up way. 

The choice of trainsition probabities $q(S^*\mid S)$ penalizes tree structures with high complexity, e.g., too many \texttt{lt()} nodes, which helps control complexity of the output. Constants in $q(S^*\mid S)$ guarantee well-definedness of the probabilities, which can be changed to favor certain transitions over others.

\subsection{Jump between Spaces of Parameters}

Another issue of proposing new structure $S^*$ is that the number of linear transformation nodes may change. Therefore the dimensionality of $\Theta$ may be different and RJMCMC (reversible jump Markov Chain Monte Carlo) proposed in \cite{green1995reversible} settles the problem well. 

After we generate $S^*$ from $S$, there are three situations.
\begin{itemize}
    \item \textbf{No Change.} When the new structure does not change the number of \texttt{lt()} nodes, the dimensionality of parameters does not change. In this case, it is sufficient to use ordinary MH step. Here the set of \texttt{lt()} nodes may change, but the sampling of new parameters is i.i.d., so we are satisfied with the MH step.
    
    \item  \textbf{Expansion.} When the number of \texttt{lt()} nodes increases, the dimensionality of $\Theta$, denoted by $p_\Theta$, increases. We may simultaneously lose some original \texttt{lt()} nodes and have more new ones. But due to the i.i.d. nature of parameters we only consider the number of all \texttt{lt()} nodes.
    
  Denote the new parameter as $\Theta^*$. According to RJMCMC, we sample auxiliary variables $U=(u_\Theta,u_n)$ where $\dim(u_\Theta)=\dim(\Theta)$, $\dim(u_n)+\dim(\Theta)=\dim(\Theta^*)$. 
    
    The hyper-parameters $U_\sigma=(\sigma_a^2,\sigma_b^2)$ are independently sampled from the inverse gamma prior, then each element of $u_\Theta$ and $u_n$ is independently sampled from $N(1,\sigma_a^2)$ or $N(0,\sigma_b^2)$ accordingly. The new parameter $\Theta^*$ along with new auxiliary variable $U^*$ is obtained by 
    \begin{small}
    \begin{align*}
    (U^*,\Theta^*,\sigma_\Theta^*) &= j_{e}(\Theta,U,U_\sigma)=j_e(\Theta,u_\Theta,u_n,U_\sigma) \\
    &= \Big(\frac{\Theta - u_\Theta}{2},\frac{\Theta + u_\Theta}{2},u_n,U_\sigma \Big),
    \end{align*}
    \end{small}
    where $U^*=\frac{\Theta - u_\Theta}{2},\quad \Theta^*=(\frac{\Theta + u_\Theta}{2},u_n),\quad \sigma_\Theta^* = U_\sigma$. Then we discard $U^*$ and get $\Theta^*,\sigma_\Theta^*$.
    
    \item \textbf{Shrinkage.} $\Theta$ shrinks when the number of \texttt{lt()} nodes decreases. Similar to the Expansion case, we may lose some \texttt{lt()} nodes and also have new ones (especially in the ReassignOperator transition), but only the dimensionality is of interest. Assume that the original parameter is $\Theta =(\Theta_0,\Theta_d)$ where $\Theta_d$ corresponds to the parameters of nodes to be dropped. Denote the new parameter as $\Theta^*$.
    
    Firstly, $U_\sigma = (\sigma_a^2,\sigma_b^2)$ are sampled independently from the inverse gamma prior. The new parameter candidate is then obtained by first sampling $U$, whose elements are independently sampled from $N(0,\sigma_a^2)$ and $N(0,\sigma_b^2)$, respectively, with $\dim(U)=\dim(\Theta_0)$. Then the new candidate $\Theta^*$ as well as the corresponding auxiliary variable $U^*$ is obtained by
    \begin{small}
    \begin{align*}
    (\sigma_\Theta^*,\Theta^*,U^*) &= j_{s}(U_\sigma,U,\Theta) = j_s(U_\sigma,U,\Theta_0,\Theta_d) \\
    &= (U_\sigma,\Theta_0 + U,\Theta_0-U,\Theta_d),
    \end{align*}
    \end{small}
    where $\sigma_\Theta^* = U_\sigma,\quad \Theta^* = \Theta_0 + U,\quad U^* = (\Theta_0 - U,\Theta_d)$.
    Then we discard $U^*$ and obtain $U^*$, $\sigma_\Theta^*$.
\end{itemize}

For simplicity, we denote the two transformations $j_e$ and $j_s$ as $j_{S,S^*}$, indicating a parameter transformation from $S$ to $S^*$, and the associated auxiliary variables are denoted as $U$ and $U^*$ respectively. Note that $\dim(\Theta)+\dim(U)=\dim(\Theta^*)+\dim(U^*)$ in both cases.

\subsection{Accepting New Candidates}
Return to the $K$-tree case. We sequentially update the $K$ trees in a way similar to \cite{chipman2010} and \cite{hastie2000}. Suppose we start from tree $(T_j^{(t)},M_{j}^{(t)},\Theta_{j}^{(t)})$, that is, the $j$-th tree of the $t$-th accepted model, and that the newly proposed structure is $(T_j^*,M_j^*,\Theta_j^*)$. Denote 
\begin{small}
\begin{align*}
&(T^{(t)},M^{(t)},\Theta^{(t)})=\{(T_i^{(t)},M_{i}^{(t)},\Theta_{i}^{(t)})\}_{i=1}^k,\\
&(T^*,M^*,\Theta^*)=\{(T_i^{*},M_{i}^{*},\Theta_{i}^{*})\}_{i=1}^k,
\end{align*}
\end{small}
where $(T_i^{*},M_{i}^{*},\Theta_{i}^{*})=(T_i^{(t)},M_{i}^{(t)},\Theta_{i}^{(t)})$ for $i\neq j$. Also let $S^*=(T^*,M^*)$, $S^{(t)}=(T^{(t)},M^{(t)})$. And $(\sigma^*)^2$ is the newly-sampled version of $(\sigma^{(t)})^2$. For simplicity, let $\Sigma^{(t)}=\big( (\sigma^{(t)})^2,\sigma_\Theta^{(t)}\big)$ and $\Sigma^*=\big((\sigma^*)^2,\sigma_\Theta^*\big)$.

If $\dim(\Theta_{i}^{(t)})=\dim(\Theta^*)$, the ordinary MH step gives the acceptance rate
\begin{small}
\begin{equation}
R=\frac{f(y\mid OLS(x,S^*,\Theta^*),\Sigma^*)f(S^*)q(S^{(t)}\mid S^*)}{f(y\mid OLS(x,S^{(t)},\Theta^{(t)}),\Sigma^{(t)}) f(S^{(t)})q(S^*\mid S^{(t)})}.
\label{eq:ratio_nonchange_lin}
\end{equation}
\end{small}
If $\dim(\Theta_{i}^{(t)})\neq \dim(\Theta^*)$, the RJMCMC method gives the acceptance rate

\begin{small}
\begin{equation}
\begin{aligned}
R &= \frac{f(y\mid OLS(x,S^*,\Theta^*),\Sigma^*)f(\Theta^*\mid S^*)q(S^{(t)}\mid S^*)}{f(y\mid OLS(x,S^{(t)},\Theta^{(t)}),\Sigma^{(t)})f(\Theta^{(t)}\mid S^{(t)}) q(S^*\mid S^{(t)})}\\
& \cdot \frac{f(S^*)p(\Sigma^*)h(U^*\mid \Theta^*,S^*,S^{(t)})}{f(S^{(t)})p(\Sigma^{(t)})h(U^{(t)}\mid \Theta^{(t)},S^{(t)},S^*)}\cdot \bigg| \frac{\partial j_{S^{(t)},S^*}(\Theta^{(t)},U^{(t)})}{\partial(\Theta^{(t)},U^{(t)})} \bigg|
\end{aligned}
\label{eq:ratio_change_lin}
\end{equation}
\end{small}

In each case, we accept the new candidate with probability $\alpha = \min \{1,R\}$ with $R$ in Equation (\ref{eq:ratio_nonchange_lin}) or (\ref{eq:ratio_change_lin}). If the new candidate is accepted, we next update the $(j+1)$-th tree starting from $(T^{(t+1)},M^{(t+1)},\Theta^{(t+1)})=(T^*,M^*,\Theta^*)$ and $\Sigma^{(t+1)}=\Sigma^*$. Otherwise we update the $(j+1)$-th tree starting at $(T^{(t)},M^{(t)},\Theta^{(t)})$ with $\Sigma^{(t)}$.

\section{Experiments}
We carry out BSR on both simulated data and real-world data.
Firstly, we compare fitness and generalization ability by comparing RMSEs on training and testing data. Secondly, we compare the complexity of the expressions generated by BSR and GP. Meanshile, we examine the robustness of the proposed BSR method by testing whether the estimated model is sensitive to the parameter K, which is the number of trees used in the linear regression. We also apply BSR on financial data to find effective 'signals'.

\subsection{Simualtion Designs}
\subsubsection{Benchmark Problems}
We set up a benchmark mathematical expression sets with six tasks presented in Equations \eqref{exp:1} to \eqref{exp:6}. We fit a  BSR model on each of the tasks. These formulas have been widely used to test other symbolic regression methods, including those based on GP, see \cite{chen2015generalisation},\cite{topchy2001faster} and \cite{chen2016improving}.
\begin{small}
\begin{eqnarray}
f_1(x_0,x_1) & = & 2.5x_0^4 - 1.3x_0^3 + 0.5x_1^2 - 1.7x_1 \label{exp:1}\\
f_2(x_0,x_1) & = & 8x_0^2+8x_1^3-15 \label{exp:2}\\
f_3(x_0,x_1) & = & 0.2x_0^3 + 0.5x_1^3 - 1.2x_1 - 0.5x_0 \label{exp:3}\\
f_4(x_0,x_1) & = & 1.5\exp(x_0) + 5\cos(x_1) \label{exp:4}\\
f_5(x_0,x_1) & = & 6.0\sin(x_0)\cos(x_1) \label{exp:5}\\
f_6(x_0,x_1) & = & 1.35x_0x_1 + 5.5\sin\{(x_0-1)(x_1-1)\} \label{exp:6}
\end{eqnarray}
\end{small}
\subsubsection{Datasets}
Simulation studies in \cite{chen2015generalisation} are adopted here. For each target formula, we have one training dataset and three testing datasets. The training set consists of 100 samples with predictors generated independently from $U[-3,3]$, and form the response variable with the corresponding formula above. We consider three different testing sets, all with size of 30. Predictors of the three testing sets are generated from $U[-3,3]$, $U[-6,6]$ and $U[3,6]$, respectively.

\subsubsection{Parameter Settings}
Note that the GP algorithm consists of two nested iterations, the inner loop for population and the outer loop for generation. Therefore, the number of trees generated by GP is $N_g \times N_p$, where $N_g$ is the number of generations and $N_p$ is the population size. We set $N_g=200$ and $N_p=100$ here, generating a total of 200,000 trees. For BSR, 100,000 trees are generated in total (in experiments, it typically consumes less than 10,000 candidates to stable results). In addition, we specify $K=2$ additive components for BSR for all tasks.
The basis function pool is $\{+,-,\times,\div, \sin, \cos, \exp, x^2, x^3\}$ for both methods. 
In order to see the stability of their performances, we run the two methods in each task for 50 times independently. 

\subsection{Simualtion Results}
\subsubsection{Accuracy and Generalization Abilities}
We use root mean square error (RMSE) to measure fitness on training data, and use RMSE on testing set to see generalization. The performances including mean and standard deviation of RMSEs are summarized in Table \ref{table:rmses}. It turns out that BSR outperforms GP in most tasks, except Equation (\eqref{exp:6}). A plausible reason is that the structure is far from linear structure, which is one of the key assumptions of BSR. 

\begin{table}[!htbp]\footnotesize
\centering
\caption{RMSEs of Both Methods}
\begin{tabular}{c|c|c|c}

\hline
  &  & \multicolumn{2}{c}{\textbf{RMSEs~(mean $\pm$ std)}} \\ \cline{3-4}
\textbf{Task}        & \textbf{Dataset} & BSR & GP  \\ \hline
\multirow{4}{*}{$f_1$} & train[-3,3] & $2.00 \pm 3.87$ &$2.71 \pm 2.43$   \\ 
& test[-3,3] &$2.04 \pm 3.27$ & $4.25 \pm 4.59$ \\
& test[-6,6] &$92.09 \pm 258.54$& $116.29 \pm 97.59$ \\
& test[3,6] &$118.53 \pm 311.57$ &  $203.31 \pm 168.34$  \\
\hline

\multirow{4}{*}{$f_2$} & train[-3,3] & $7.30 \pm 10.19$ & $3.56 \pm 5.79$ \\ 
& test[-3,3] & $6.84 \pm 10.10$ & $2.92 \pm 4.41$  \\
& test[-6,6] & $95.33 \pm 145.31$ & $121.41 \pm 126.19$  \\
& test[3,6] & $128.27 \pm 221.73$ & $174.01 \pm 173.71$  \\
\hline

\multirow{4}{*}{$f_3$} & train[-3,3] & $0.19 \pm 0.16$ & $0.63 \pm 0.33$ \\ 
& test[-3,3] & $0.21 \pm 0.20$ & $0.60 \pm 0.35$  \\
& test[-6,6] & $9.38 \pm 9.08$ & $28.97 \pm 20.68$  \\
& test[3,6] & $15.19 \pm 32.24$ & $34.08 \pm 25.41$ \\
\hline

\multirow{4}{*}{$f_4$} & train[-3,3] & $0.14 \pm 0.56$ &$0.72 \pm 1.01$  \\ 
& test[-3,3] &$0.16 \pm 0.62$&  $0.84 \pm 1.12$  \\
& test[-6,6] &$6.96 \pm 19.44$ & $24.62 \pm 29.66$  \\
& test[3,6] &$12.06 \pm 38.27$ & $31.74 \pm 36.77$ \\
\hline

\multirow{4}{*}{$f_5$} & train[-3,3] & $0.68 \pm 1.14$ & $0.78 \pm 0.96$ \\ 
& test[-3,3] & $0.66 \pm 1.13$ & $0.72 \pm 0.83$ \\
& test[-6,6] & $1.09 \pm 2.39$ & $1.58 \pm 1.55$ \\
& test[3,6] & $1.41 \pm 3.57$ & $4.49 \pm 5.07$ \\
\hline

\multirow{4}{*}{$f_6$} & train[-3,3] & $3.99 \pm 0.71$ & $3.17 \pm 0.79$ \\ 
& test[-3,3] &$4.63 \pm 0.62$ &  $3.70 \pm 0.93$  \\
& test[-6,6] &$12.22 \pm 8.46$ & $5.13 \pm 1.91$  \\
& test[3,6] &$14.44 \pm 10.39$& $11.09 \pm 12.58$  \\
\hline

\end{tabular}
\label{table:rmses}
\end{table}

\subsubsection{Complexity of Expressions}
One of the most important aim for BSR is to improve interpretability by restricting the formula to a concise and readable form. Specifically, we introduce an additive symbolic tree structure for BSR model.   

To check if BSR achieves this aim, we summarize the complexity of the output from BSR and GP in Table \ref{table:num_nodes}, namely the means and standard deviations of number of nodes in each tree in the 50 replications.

\begin{table}[H]\footnotesize
\centering
\caption{Complexity of Expressions}
\begin{tabular}{c|c|c}

\hline
  &  \multicolumn{2}{c}{\textbf{Number of Nodes~(mean $\pm$ std)}} \\ \cline{2-3}
\textbf{Task}        & BSR & GP  \\ \hline
$f_1$ &  $\mathbf{22.16 \pm 7.44}$ &$40.85 \pm 21.34$  \\
\hline

$f_2$ &  $\mathbf{12.25 \pm 11.41}$ & $54.51 \pm 38.89$ \\
\hline

$f_3$ &  $27.23\pm 10.61$ &$22.88 \pm 8.62$ \\
\hline

$f_4$ &  $\mathbf{13.64 \pm 12.50}$  &$22.80 \pm 8.82$  \\ 
\hline

$f_5$ &  $31.28 \pm 9.13$ & $19.80 \pm 10.28$  \\
\hline

$f_6$ &  $\mathbf{20.08 \pm 4.78}$ &$21.18 \pm 25.73$  \\
\hline

\end{tabular}
\label{table:num_nodes}
\end{table}

According to Table \ref{table:num_nodes}, the number of nodes on trees generated by BSR is significantly less that those generated by GP, leading to more concise and readable expressions. Table \ref{table:expressions} lists some typical expressions output from BSR and GP, where only two cases are exhibited due to limitations of paper length, leaving others to appendix. It turns out that expressions estimated by BSR are generally closer to the ground truth and they are shorter and more comprehensible. The simulation study here verifies that, in favourable scenarios, BSR reaches its aim and shows its advantage in both prediction accuracy and interpretability.  

\begin{figure}[H]
    \centering
    \subfigure[RMSEs in training $f_1$]{
        \includegraphics[width=0.21\textwidth]{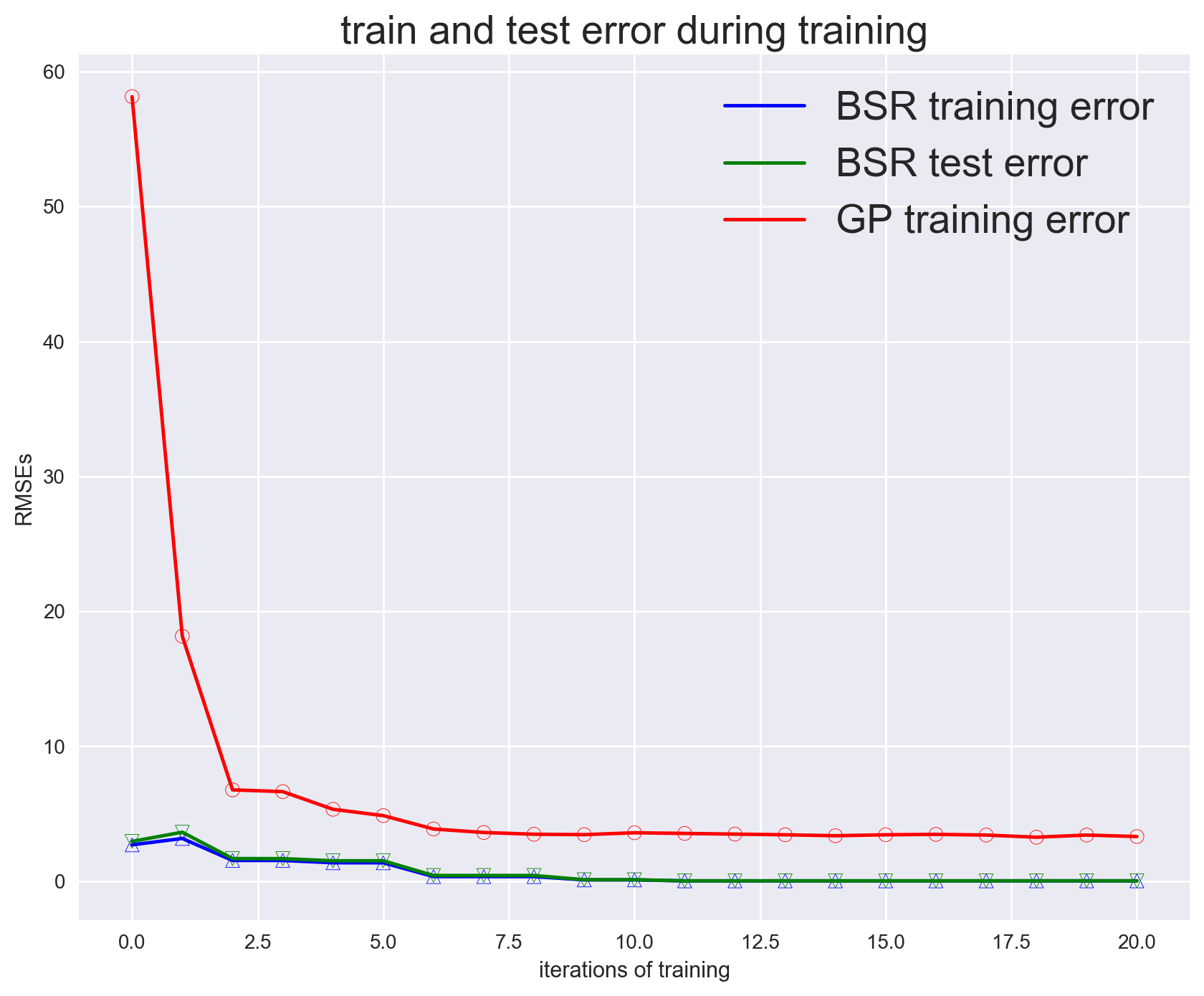}
    }
    \subfigure[Complexity in training $f_1$]{
        \includegraphics[width=0.21\textwidth]{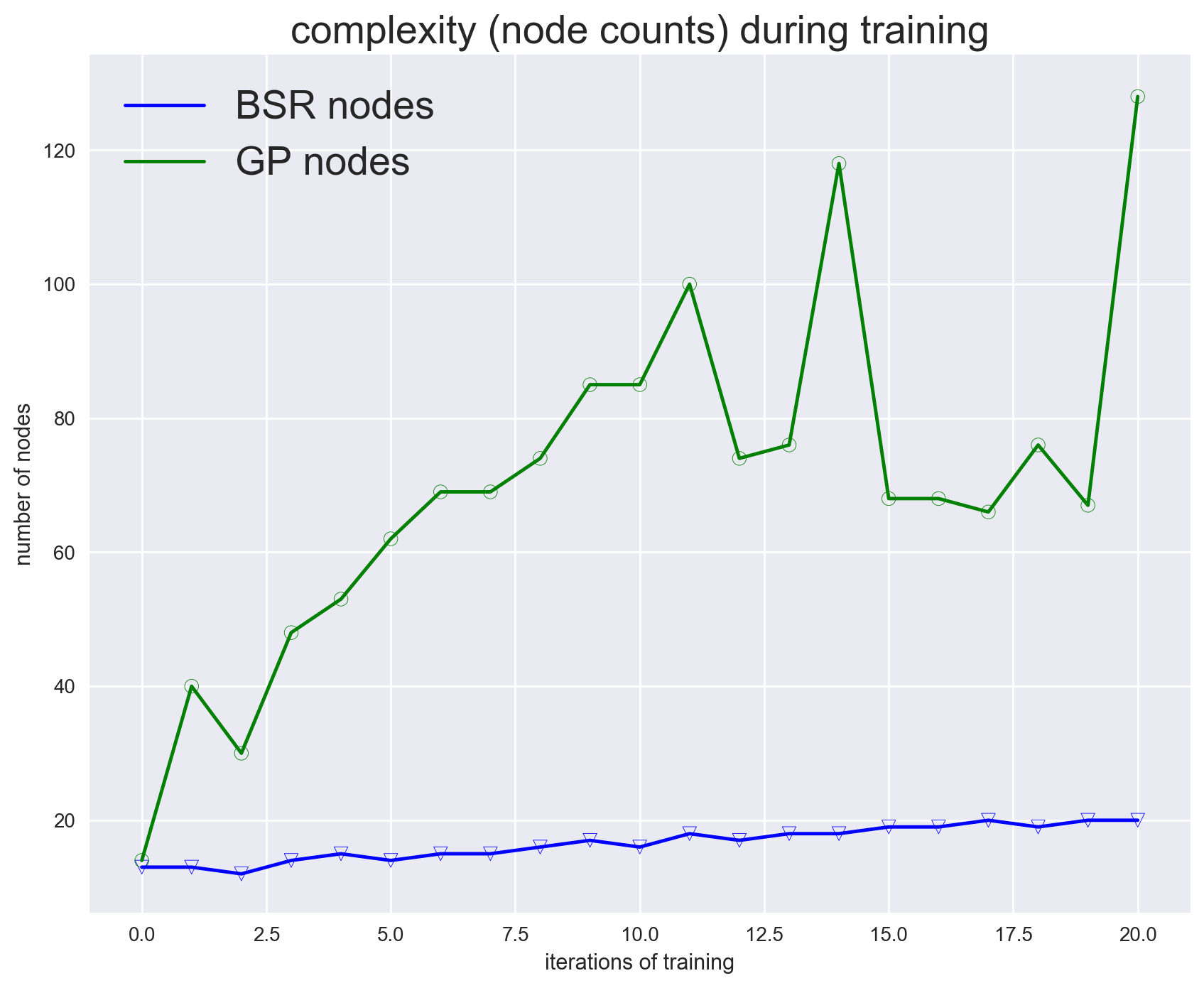}
    }
    \subfigure[RMSEs in training $f_2$]{
        \includegraphics[width=0.21\textwidth]{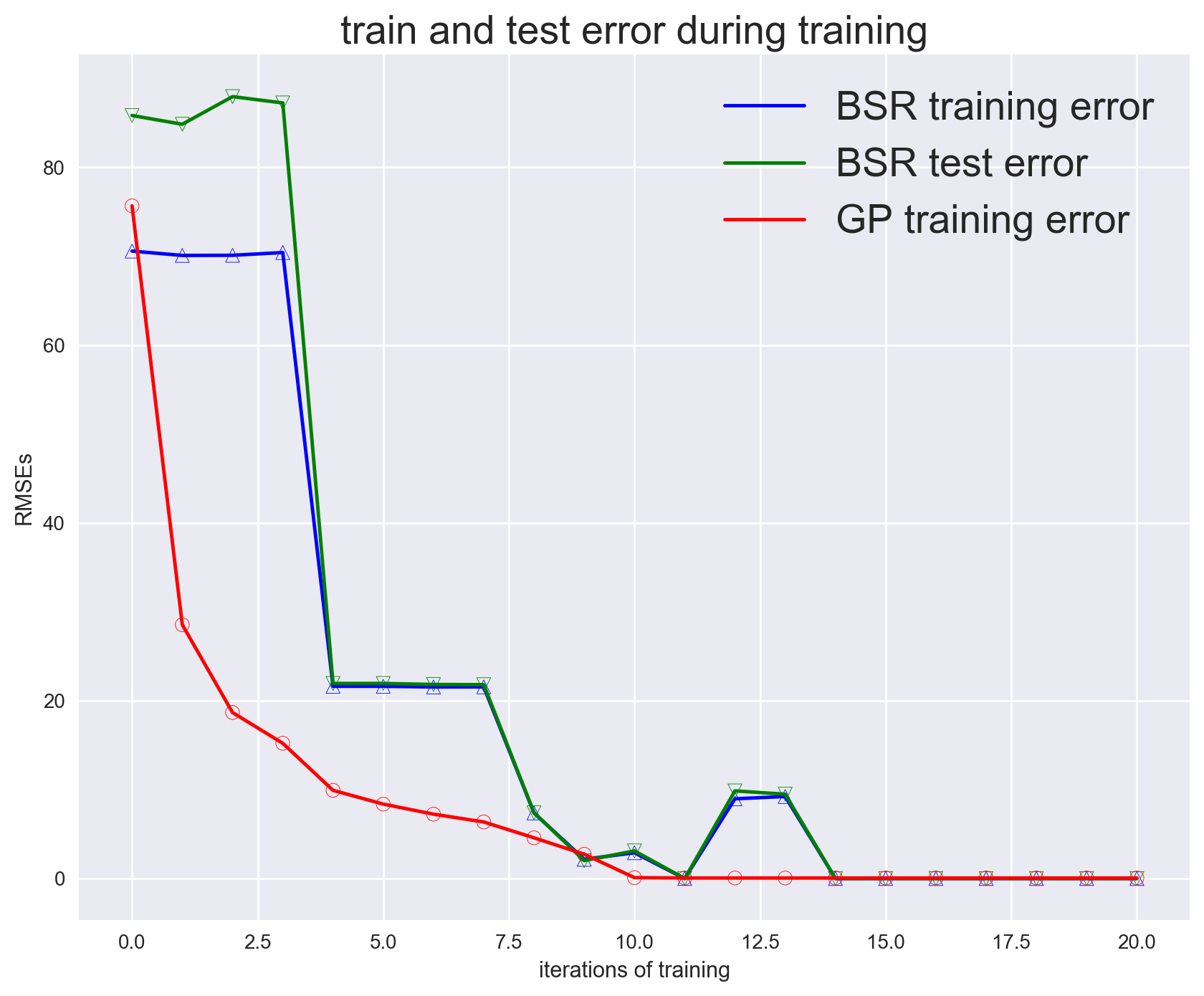}
    }
    \subfigure[Complexity in training $f_2$]{
        \includegraphics[width=0.21\textwidth]{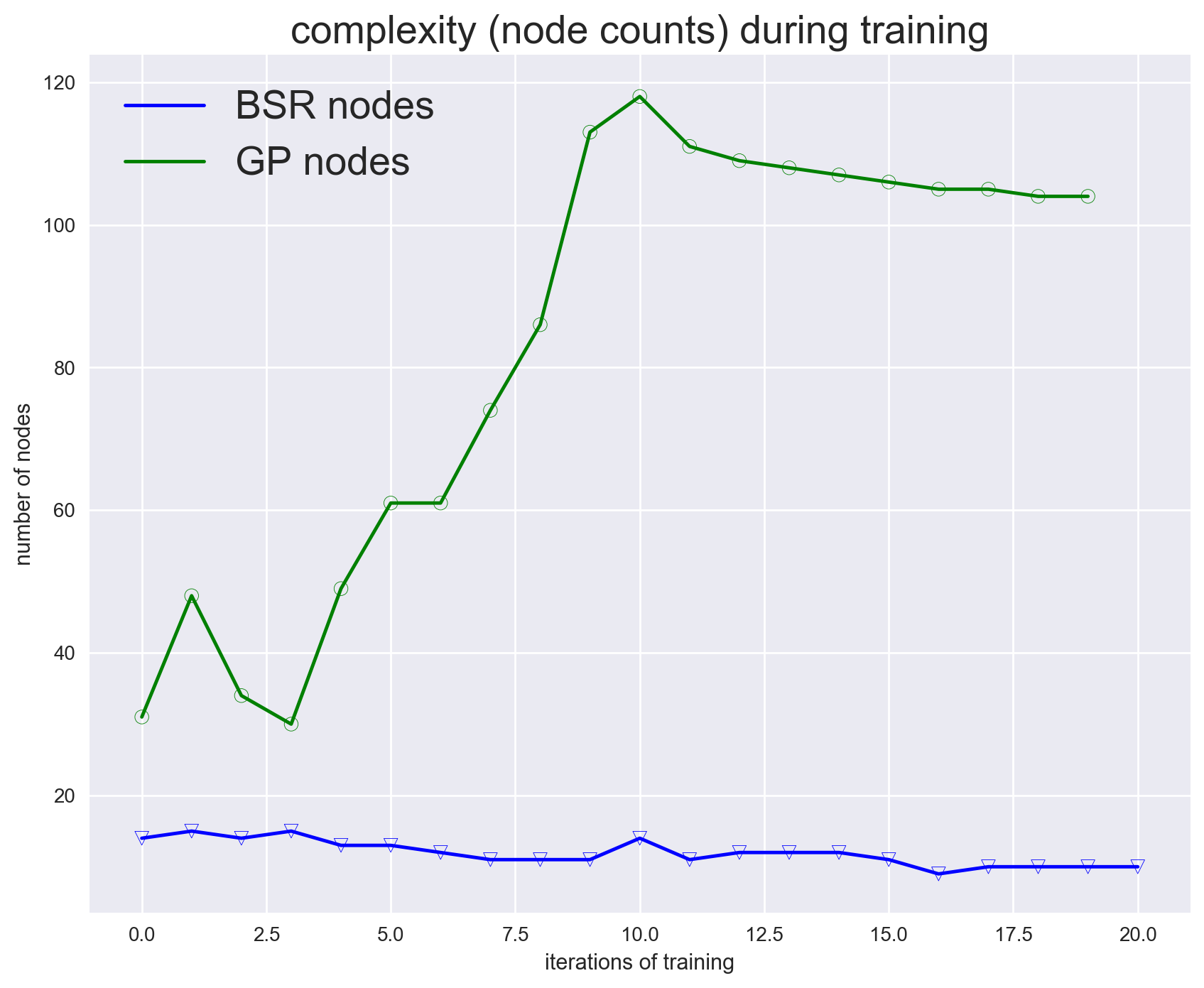}
    }

    \caption{RMSEs and Complexities during training}
    \label{fig:exp_perf} 
\end{figure} 

To further illustrate the performance of BSR versus GP in typical training processes, we plot the RMSEs of training data and testing dataset on $[-3,3]$ for BSR at every acceptence during the training (See Figure \ref{fig:exp_perf}). We also include the training RMSE of best individual at generations of GP, which are evenly-paced to match the number of records. Also we compare the complexity of models, evaluated by number of nodes in the tree. Due to limitation of paper length, we only exhibit results for $f_1$ and $f_2$, leaving others to appendix. Figure \ref{fig:exp_perf} shows that BSR reduces both training and testing RMSE during the training process, with less complex outputs compared to GP.

\begin{table}[H]\scriptsize
\centering
\renewcommand\arraystretch{1}
\caption{Typical Expressions}
\begin{tabular}{c|c|l}

\hline
  \textbf{Task}   &  \multicolumn{2}{c}{\textbf{ Expressions}} \\ \hline

\multirow{4}{*}{$f_1$} & Truth & $f_1 = 2.5x_0^4 - 1.3x_0^3 + 0.5x_1^2 - 1.7x_1$ \\\cline{2-3}
&  \multirow{6}{*}{GP} & $y=((exp((\frac{-x_0}{0.80}+0.81))-(((\sin((0.80x_0)^2)$\\
&                      & $-\cos(x_1)^6)+\sin((0.80x_0)^2))+\cos(x_1)))$\\
&                      & $-(((\frac{x_0}{-0.80})+((((\frac{x_0}{-0.80})+\cos(x_1))+((\sin((0.71x_0)^2)$\\
&                      & $-((\sin(((0.71x_0))^2)-0.77))^2)+1.0))+x_1))$\\
&                      & $+(0.76+x_1)))+(((\frac{x_0^2}{0.78}))^2+\frac{x_0^2}{0.80})$ \\\cline{2-3}
& \multirow{2}{*}{BSR} &$y=(-0.02)+(-1.30)[x_0^3+1.30x_1+0.09]$\\
& & $+(0.49)[5.05x_0^4+x_1^2+0.31]$ \\\hline

\multirow{4}{*}{$f_2$} & Truth & $f_2 = 8x_0^2+8x_1^3-15$ \\\cline{2-3}
&  \multirow{3}{*}{GP} & $y=(exp(1.82)x_1^3)+5.26(x_0^2-(\cos((0.90x_0))$\\
&                      & $*(exp(0.187)+\cos((x_0^2\cos(0.75))))))$\\
&                      & $+(x_1-0.77)^3+exp(x_1-0.38) (x_1-0.38)$ \\\cline{2-3}
& \multirow{2}{*}{BSR} &$y=(-0.02)+(-1.38)[-7.56x_0^2+2.85]$\\
&                      & $+(8.00)[-0.30x_0^2+x_1^3-1.38] $\\\hline
\end{tabular}
\label{table:expressions}
\end{table}

\begin{table}[!htbp]
\footnotesize
\centering
\caption{RMSEs for different K}
\begin{tabular}{c|ccc}

\hline
  & \multicolumn{3}{c}{\textbf{RMSE~(mean $\pm$ std)}} \\ \cline{2-4}
\textbf{Task}        & K=2 & K=4 & K=8 \\ \hline

$f_1$      & $2.04\pm3.27$ & $2.86\pm5.04$ & $0.64\pm2.46$ \\ \hline
$f_2$      & $6.84\pm10.10$ & $0.02\pm0.03$ & $0.03\pm0.1$ \\ \hline
$f_3$      & $0.21\pm0.20$ & $0.06\pm0.03$ & $0.03\pm0.02$ \\ \hline
$f_4$      & $0.16\pm0.62$ & $0.03\pm0.06$ & $0.01\pm0.01$ \\ \hline
$f_5$      & $0.66\pm1.13$ & $0.29\pm0.80$ & $0.42\pm0.94$ \\ \hline
$f_6$      & $4.63\pm0.62$ & $4.00\pm0.34$ & $5.28\pm4.38$ \\ \hline

\end{tabular}
\label{table:K}
\end{table}

\subsubsection{Sensitivity to the Number of Components $K$}
The number of additive components $K$ is an important hyper-parameter in BSR model and it is interesting to study if the optimal expression selected by BSR is sensitive to the choice of $K$. To check this, we summarize the average RMSEs on testing set $[-3,3]$ out of 50 replications in Table \ref{table:K}.

It turns out that RMSEs of these tasks are smaller as $K$ grows, but the improvement of performance is not significant when $K$ is large enough. It is interesting to see that even if $K$ is set to be smaller than ground truth, BSR can automatically find an approximately equivalent additive component structure in some single trees. On the other hand, when $K$ is significantly larger than what it should be, BSR automatically ''discards'' the redundant trees by producing small coefficients in linear combination, making them similar to white noise. 

\subsection{Experiments on Real World Data}

In the quantitative finance industry, the most important task is to find 'alpha signals' effective in predicting returns of financial securities such as stocks, futures and other derivatives. These signals can be expressed as mathematical formulas such as classic factors \cite{fama1996multifactor}. However, mining signals manually is extremely inefficient, and search directions are usually biased by human knowledge. BSR provides an automatic way to select effective signals. We apply BSR on financial data to this end.

\subsubsection{Datasets and Experimental Setting}
We collected the CSI 300 INDEX data, which includes time series about daily prices from 2004 to 2019. Each record consists of five attributes: open price, high price, low price, and close price, which corresponds to the trading date. A total of 3586 records are collected in this study.

In our experiments, we define the label as the sign of the return derived from the close price in Equation (\ref{eq:return_def}). The other four attributes are predictors. In Equation (\ref{eq:return_def}), $\mbox{Close\_Price}(t)$ is the close price on the $t$th trading day, and $\mbox{Close\_Price}(t+1)$ means the close price on the next trading day. We set the sequence from 2004 to 2016 as the training set, and those from 2017 to 2019 as the testing set.

\begin{small}
\begin{equation}
\mbox{Return}(t) = \frac{\mbox{Close\_Price}(t+1)-\mbox{Close\_Price(t)}}{\mbox{Close\_Price}(t)}
\label{eq:return_def}
\end{equation}
\end{small}

We set the number of trees generated by BSR as 10,000 and the number of additive components as 2. The basis function pool is set as $\{+,-,\times,\div,\exp, x^2, x^3\}$

\subsubsection{Experiment Result}
To check if BSR can generate effective factors, we run the task for 200 times independently. A single factor with an accuracy larger than 0.5 in both the train set and the test set is considered useful. Finally, 15 expressions in the posterior modes meet that requirement. We only exhibit one expression in Expression (\ref{eq:expression_example}) due to limitations of paper length, leaving others to appendix. Expression (\ref{eq:expression_example}) achieves an accuracy of 0.539 in the train set and an accuracy of 0.518 in the test set. An intuitive explanation for Expression (\ref{eq:expression_example}) is that the relative sizes of open price and low price can predict the return, and if the open price is much higher than the low price, the return will be more likely to be positive than negative.

\begin{small}
\begin{equation}
2.9*10^{-4} - 1.2*10^{-3}*\frac{1}{open^2}+1.9*10^{-3}\frac{1}{low^2}
\label{eq:expression_example}
\end{equation}
\end{small}

\section{Conclusions and Future Research}
This paper proposes a new symbolic regression method based on Bayesian statistics framework. Compared with traditional GP, the proposed method exhibits its advantage in better model interpretability, simpler way to incorporate prior knowledge and more cost-effective memory usage etc. 

In the future, we are to continue to improve BSR in several ways. For example, we will study new MCMC algorithms to improve the search and sampling efficiency; we will study a dynamic empirical bayes method to optimize hyper-parameters in BSR; we will also study how to extend the proposed algorithm for distributed computing to improve computational efficiency. 

\clearpage
\bibliographystyle{aaai}
\bibliography{reference}

\section{Appendix}

\subsection{Pseudo-codes of BSR}
We sum up the BSR algorithm as follows.
\begin{small}
\begin{breakablealgorithm}
\caption{pseudo-codes of MCMC-based Symbolic Regression for linearly-mixed signals}
\begin{algorithmic}[1]
    \Require Datapoints $x_1,\dots,x_n$, labels $y=(y_1,\dots,y_n)$; 
    number of components $K$, number of acceptance $N$; transition kernel (proposal) $q(\cdot\mid \cdot)$, prior distributions $p(T,M,\Theta)$, likelihood function $f(y\mid OLS(S,\Theta,x))$;
    \Ensure A chain of accepted models $(T^{(t)},M^{(t)},\Theta^{(t)})$;
    
    \State From prior $p(T,M,\Theta)$, generate independently $K$ tree models (structures and parameters) $(T_i^{(1)},M_i^{(1)},\Theta_i^{(1)})$, $i=1,\dots,K$;
    \State Calculate linear regression coefficients $\beta^{(1)}$ from datapoints $x_i$, labels $y_i$ and models $(T_i^{(1)},M_i^{(1)},\Theta_i^{(1)})$, $i=1,\dots,n$ using OLS;
    \State Number of accepted models $m=1$;
    \While{$m<N$}
    \For{$i=1 \to K$}
    \State Propose $S_i^*=(T_i^*,M_i^*)$ by sampling  $S_i^*\mid S_i^{(m)} \sim q(\cdot;S_i^{(m)})$;
    \If {$dim(\Theta_i^*)\neq dim(\Theta_i^{(m)})$}
    \State Sample $U_i^{(m)}\sim h(U_i^{(m)}\mid \Theta_i^{(m)},S_i^{(m)},S_i^*)$;
    \State Obtain $(U_i^*,\Theta_i^*)=j_{S_i^{(m)},S_i^*}(\Theta_i^{(m)},U_i^{(m)})$;
    \State Calculate linear regression coefficients $\beta^*$ from datapoints $x_i$, labels $y_i$ and models $(T^*,M^*,\Theta^*)$ using OLS;
    \State Calculate the ratio $R$ in Equation (2);
    \Else
    \State Directly sample $\Theta_i^* \sim p(\cdot\mid S_i^*)$;
    \State Calculate coefficients $\beta^*$ from $x_i$, $y_i$, $i=1,\dots,n$ and models $(T^{(m)},M^{(m)},\Theta^{(m)})$ using OLS;
    \State Calculate the ratio $R$ in Equation (1);
    \EndIf
    \State $\alpha \gets \min(1,R)$;
    \State Sample $u\sim U(0,1)$;
    \If {$u<\alpha$}
        \For {$j=1 \to K$}
            \If {$j=i$}
                \State $S_j^{(m+1)}\gets S_j^*$, $\Theta_j^{(m+1)}\gets \Theta_j^*$;
            \Else 
                \State $S_j^{(m+1)}\gets S_j^{(m)}$, $\Theta_j^{(m+1)}\gets \Theta_j^{(m)}$;
            \EndIf
        \EndFor
        \State $\beta^{(m+1)}\gets \beta^*$;
        \State $m \gets m+1$;
    \EndIf
    \EndFor

    \EndWhile

        \end{algorithmic}
    \end{breakablealgorithm}
\end{small}

\subsection{Simulation results}
\subsubsection{Performance visualizations}
Figures on accuracy and complexity of BSR and GP on simulated data which are not included in the paper are summarized below.

\begin{figure}[H]
    \centering
    \subfigure[RMSEs in training $f_3$]{
        \includegraphics[width=0.21\textwidth]{f1error.png}
    }
    \subfigure[Complexity in training $f_3$]{
        \includegraphics[width=0.21\textwidth]{f1nodes.png}
    }
    \subfigure[RMSEs in training $f_3$]{
        \includegraphics[width=0.21\textwidth]{f2error.png}
    }
    \subfigure[Complexity in training $f_3$]{
        \includegraphics[width=0.21\textwidth]{f2nodes.png}
    }
    \subfigure[RMSEs in training $f_3$]{
        \includegraphics[width=0.21\textwidth]{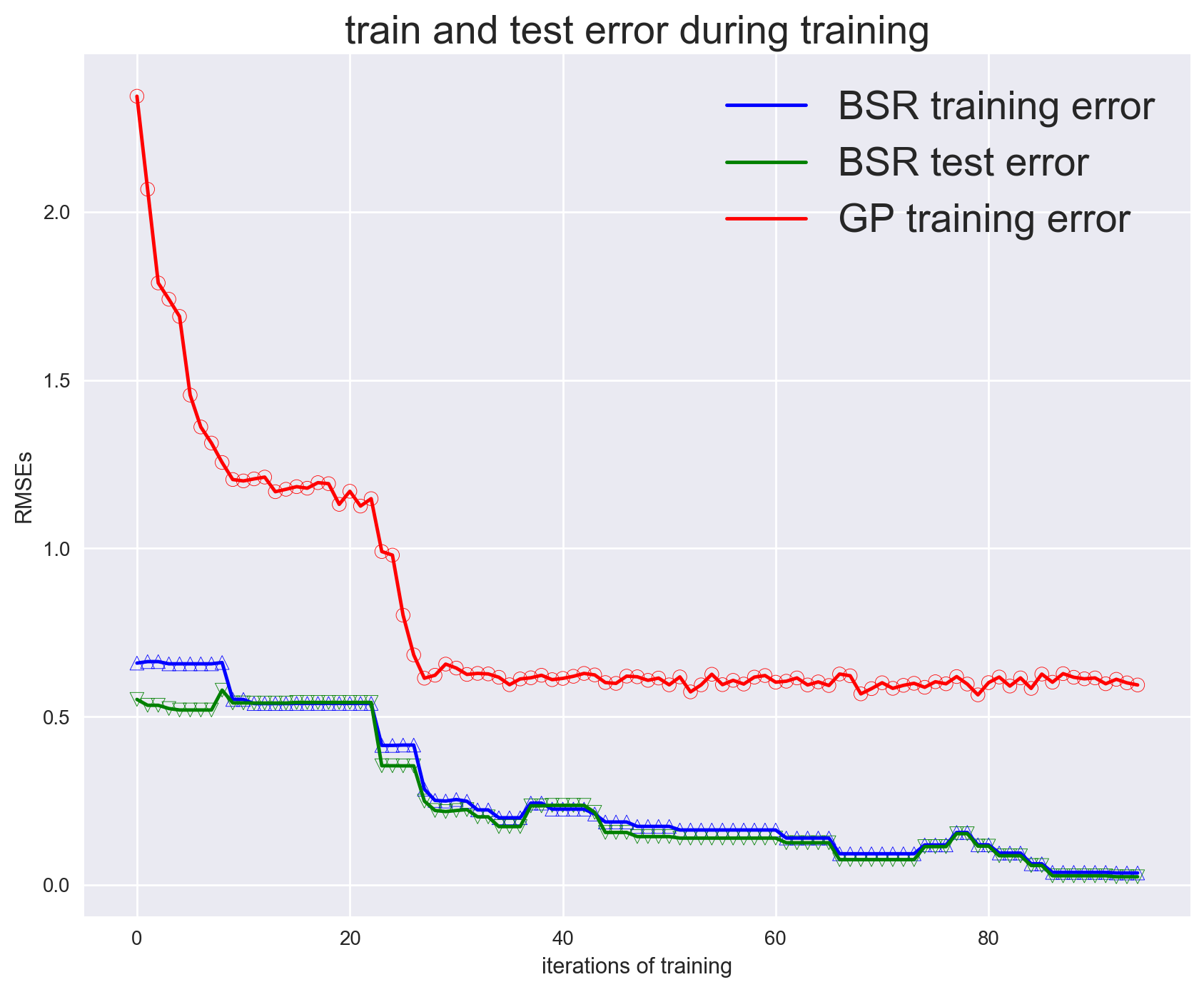}
    }
    \subfigure[Complexity in training $f_3$]{
        \includegraphics[width=0.21\textwidth]{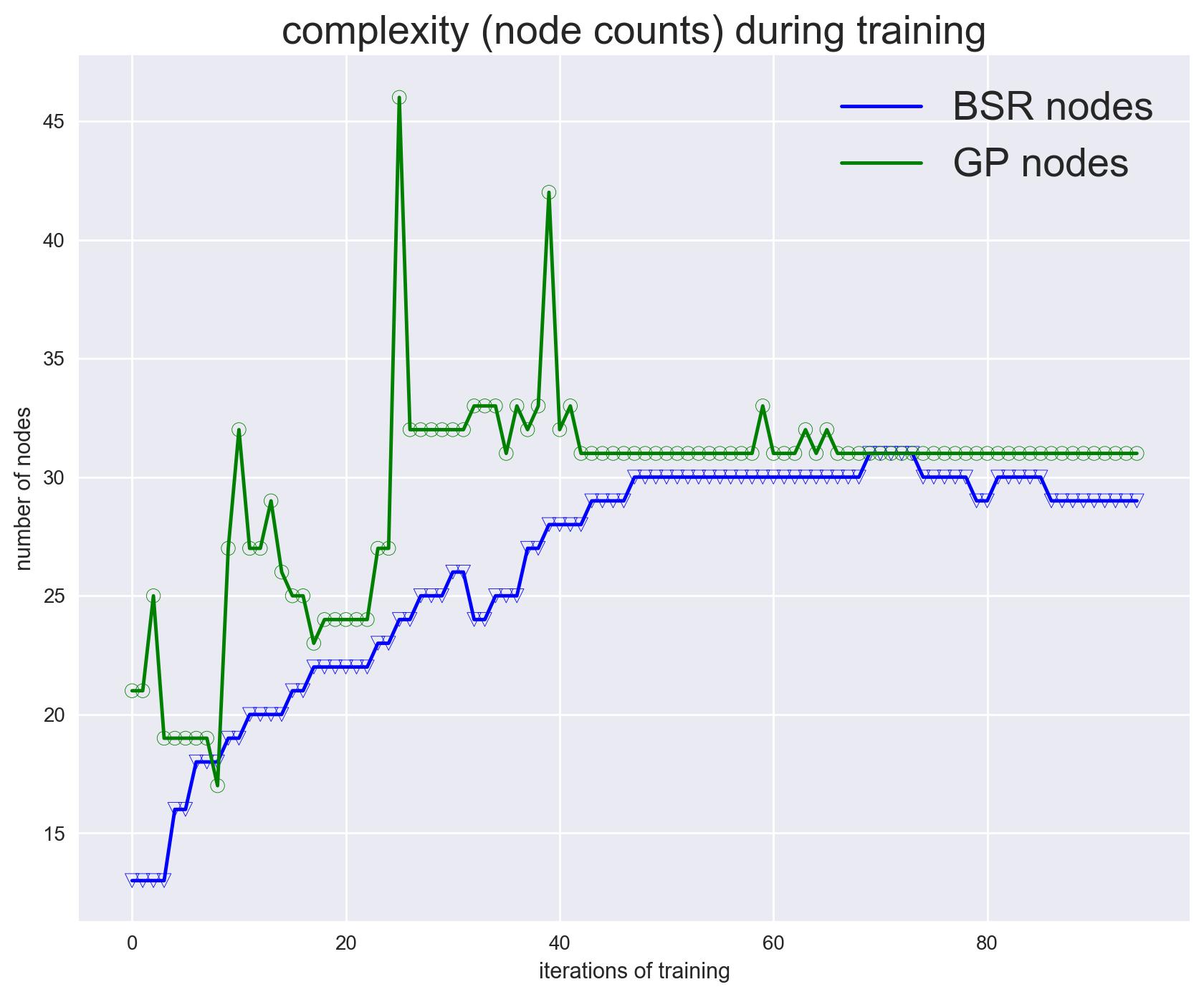}
    }
    \subfigure[RMSEs in training $f_4$]{
        \includegraphics[width=0.21\textwidth]{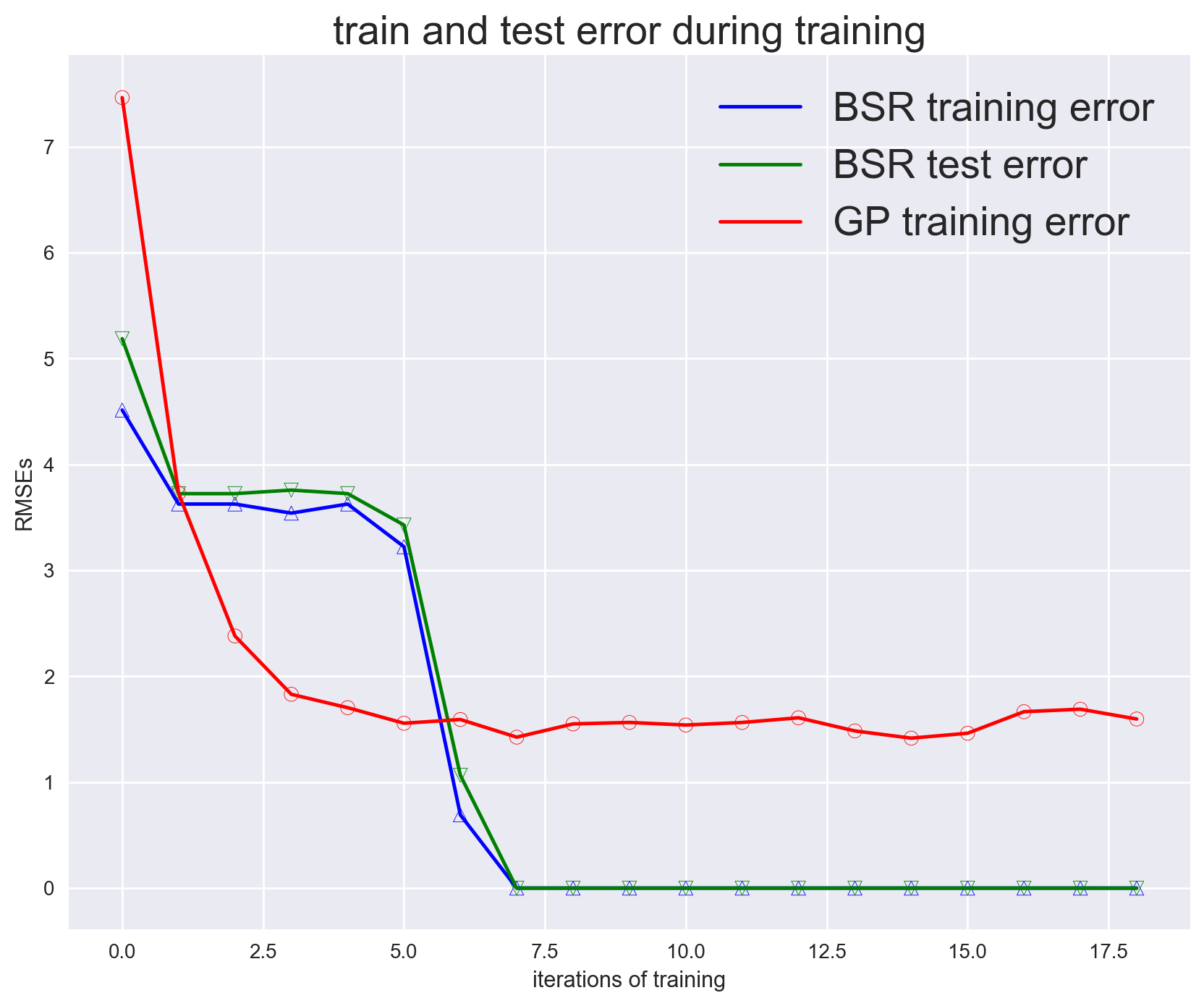}
    }
    \subfigure[Complexity in training $f_4$]{
        \includegraphics[width=0.21\textwidth]{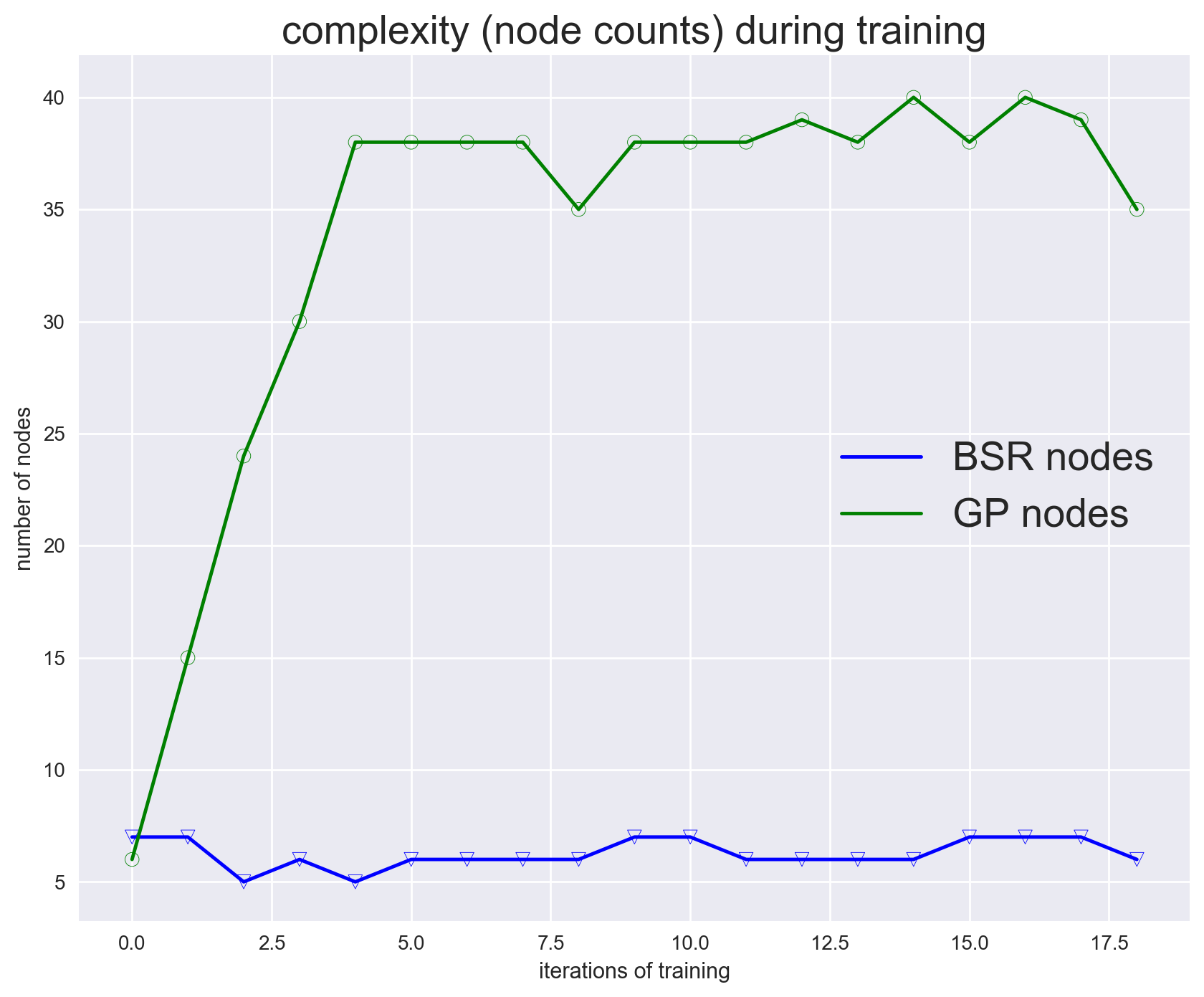}
    }
    \subfigure[RMSEs in training $f_5$]{
        \includegraphics[width=0.21\textwidth]{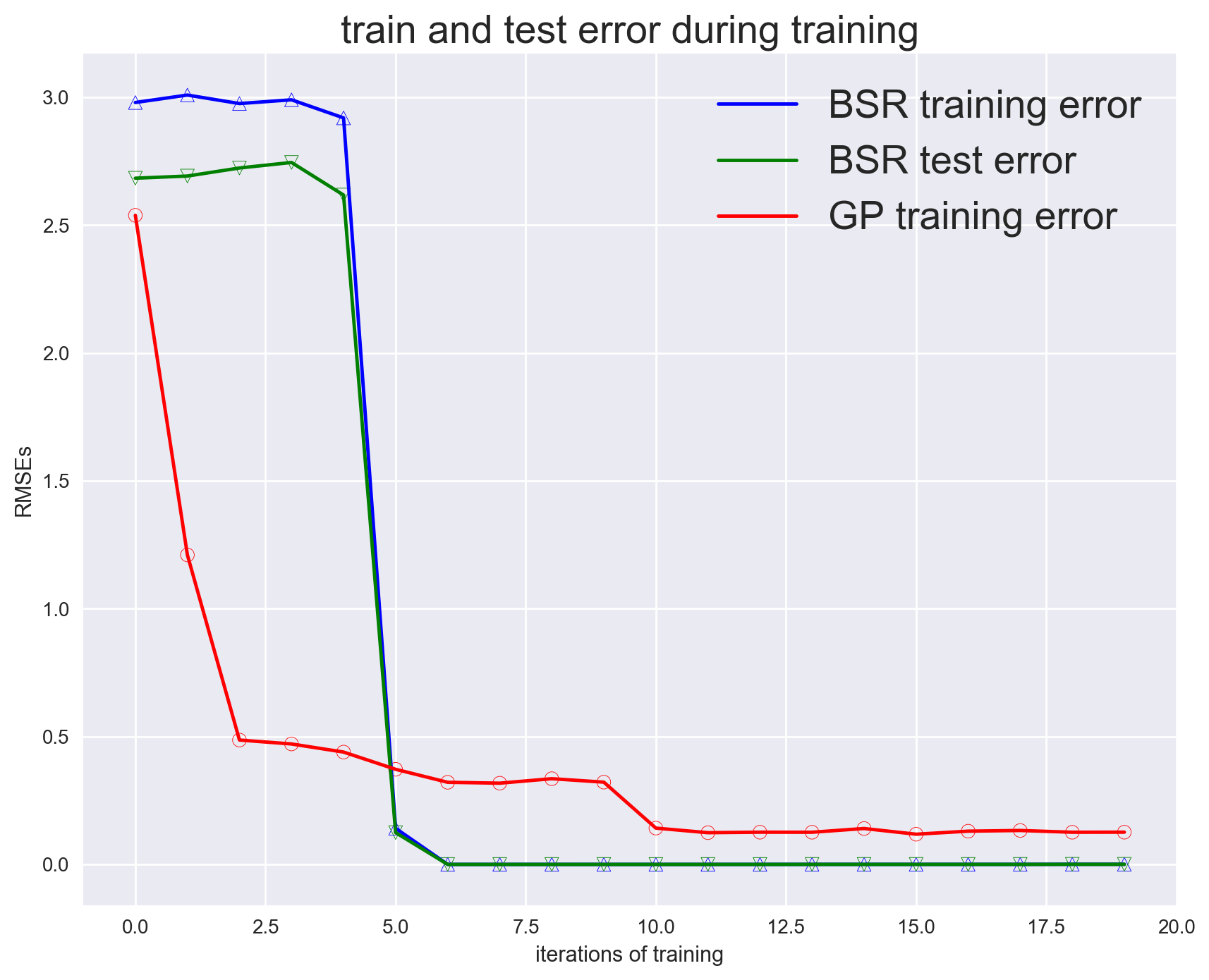}
    }
    \subfigure[Complexity in training $f_5$]{
        \includegraphics[width=0.21\textwidth]{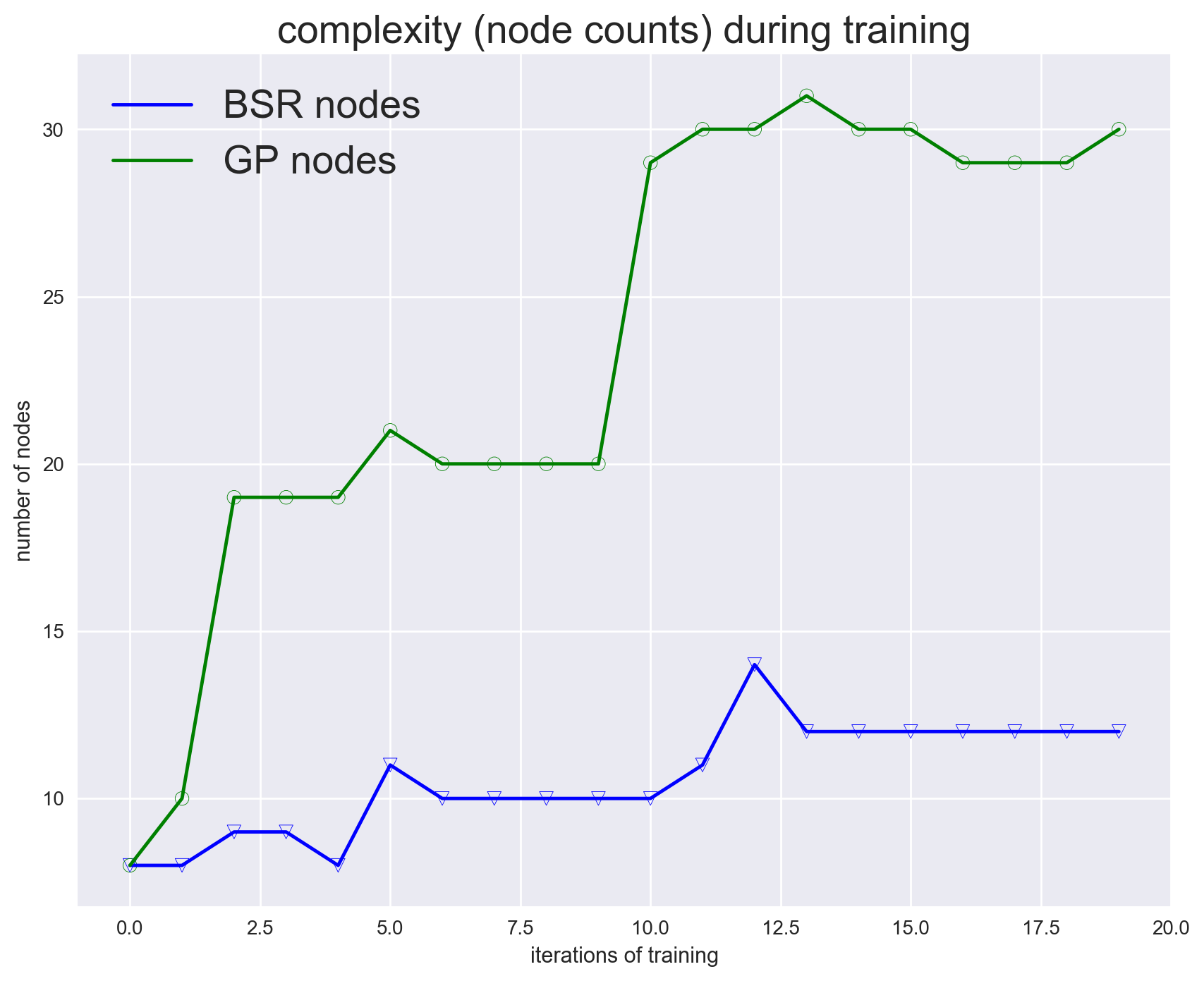}
    }
    \subfigure[RMSEs in training $f_6$]{
        \includegraphics[width=0.21\textwidth]{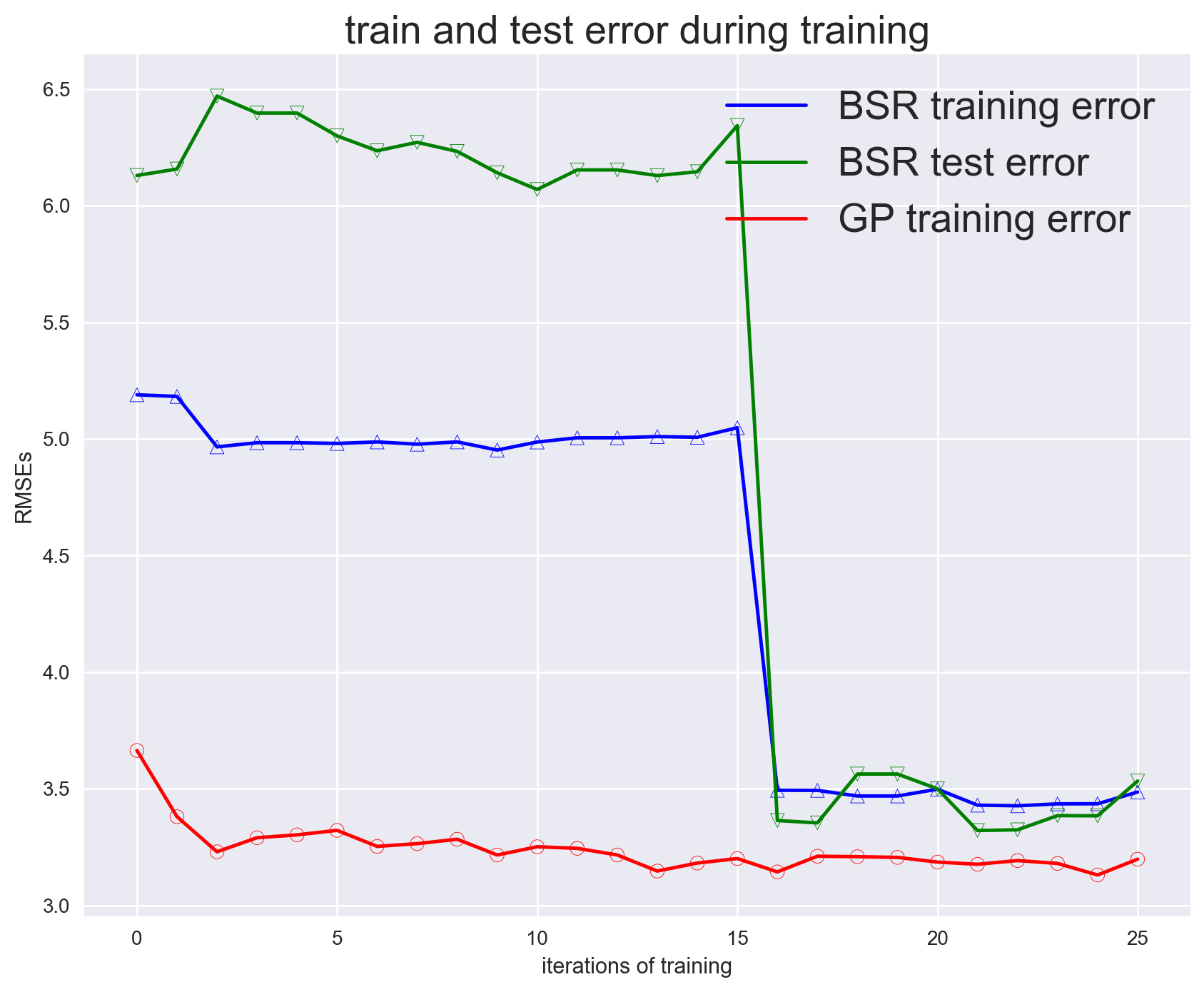}
    }
    \subfigure[Complexity in training $f_6$]{
        \includegraphics[width=0.21\textwidth]{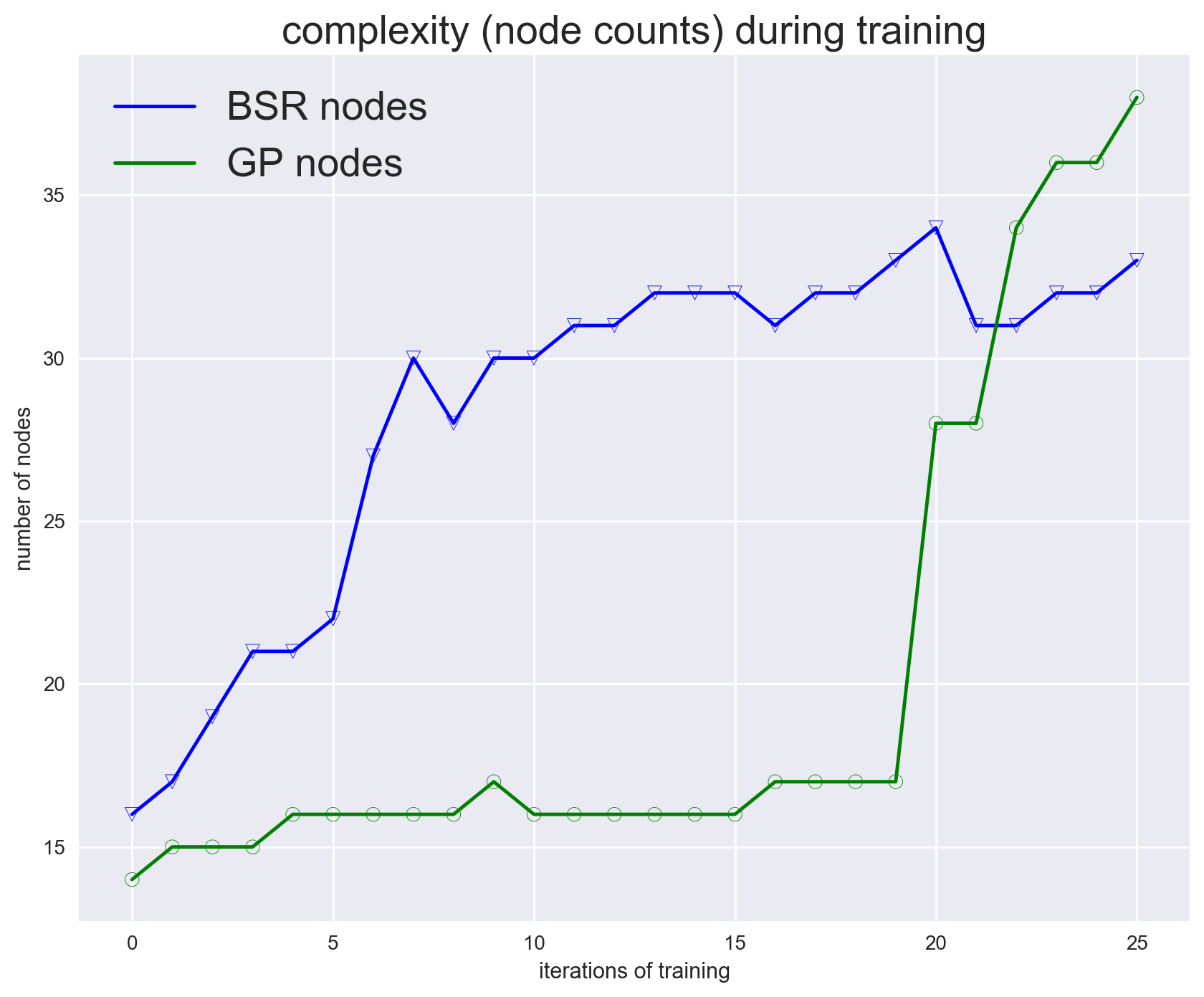}
    }
\end{figure} 

\subsubsection{Typical expressions}
Typical expressions produced by BSR and GP in the simulation studies are summarized below.

\begin{table}[H]\scriptsize
\centering
\renewcommand\arraystretch{1}
\caption{Typical Expressions}
\begin{tabular}{c|c|l}

\hline
  \textbf{Task}   &  \multicolumn{2}{c}{\textbf{ Expressions}} \\ \hline

\multirow{4}{*}{$f_1$} & Truth & $f_1 = 2.5x_0^4 - 1.3x_0^3 + 0.5x_1^2 - 1.7x_1$ \\\cline{2-3}
&  \multirow{6}{*}{GP} & $y=((exp((\frac{-x_0}{0.80}+0.81))-(((\sin((0.80x_0)^2)$\\
&                      & $-\cos(x_1)^6)+\sin((0.80x_0)^2))+\cos(x_1)))$\\
&                      & $-(((\frac{x_0}{-0.80})+((((\frac{x_0}{-0.80})$\\
&                      & $+\cos(x_1))+((\sin((0.71x_0)^2)$\\
&                      & $-((\sin(((0.71x_0))^2)-0.77))^2)+1.0))+x_1))$\\
&                      & $+(0.76+x_1)))+(((\frac{x_0^2}{0.78}))^2+\frac{x_0^2}{0.80})$ \\\cline{2-3}
& \multirow{2}{*}{BSR} &$y=(-0.02)+(-1.30)[x_0^3+1.30x_1+0.09]$\\
& & $+(0.49)[5.05x_0^4+x_1^2+0.31]$ \\\hline

\multirow{4}{*}{$f_2$} & Truth & $f_2 = 8x_0^2+8x_1^3-15$ \\\cline{2-3}
&  \multirow{3}{*}{GP} & $y=(exp(1.82)x_1^3)+5.26(x_0^2-(\cos((0.90x_0))$\\
&                      & $*(exp(0.187)+\cos((x_0^2\cos(0.75))))))$\\
&                      & $+(x_1-0.77)^3+exp(x_1-0.38) (x_1-0.38)$ \\\cline{2-3}
& \multirow{2}{*}{BSR} &$y=(-0.02)+(-1.38)[-7.56x_0^2+2.85]$\\
&                      & $+(8.00)[-0.30x_0^2+x_1^3-1.38] $\\\hline

\multirow{4}{*}{$f_3$} & Truth & $f_3 = 0.2x_0^3 + 0.5x_1^3-1.2x_1-0.5x_0$ \\\cline{2-3}
&  \multirow{3}{*}{GP} & $y=(4x_1-\sin(1.32x_1)-0.69$\\
&                      & $-(\sin(\sin(1.32x_1)/0.50) / 0.76))$\\
&                      & $-\sin(x_0)-\sin(\sin(\sin((\cos(x_1)+x_1))))$ \\\cline{2-3}
& \multirow{2}{*}{BSR} &$y=(0.04)+(-0.30)[-0.67x_0^3+4.27]$\\
& & $+(-0.21)[-2.45x_1^3+2.45x_0+x_1-0.93]$\\\hline

\multirow{4}{*}{$f_4$} & Truth & $f_4 = 1.5\exp(x_0) + 5\cos(x_1)$ \\\cline{2-3}
&  \multirow{4}{*}{GP} & $y=(((((exp(\cos(x_0))+0.59 + x_0) $\\
&                      & $+ exp(x_0))-\cos(exp(\cos(x_1))))$\\
&                      & $-\cos(exp(\cos(\sin(x_1)x_0))))$\\
&                      & $-x_1^2+x_0^2)$ \\\cline{2-3}
& \multirow{2}{*}{BSR} &$y = (-0.01)+(0.28)[17.74\cos(x_1)+0.45]$\\
&                      & $+(0.24)[6.26exp(x_0)-0.47]$\\\hline

\multirow{4}{*}{$f_5$} & Truth & $f_5 = 6.0\sin(x_0)\cos(x_1)$ \\\cline{2-3}
&  \multirow{2}{*}{GP} & $y=0.77exp(exp(\sin(\sin(\cos(0.73x_0)))))$\\
&                   & $*x_0\cos(x_1)$\\\cline{2-3}
& \multirow{3}{*}{BSR} &$y=(-7.06*10^{-9}) + (6.00)[\sin(x_0)\cos(x_1)]$\\
& & $+(2.66*10^{-9})[\sin(\frac{0.34}{\sin^2(x_1)}$\\
& & $-0.93exp(x_0+x_1)-0.95)]$\\\hline

\multirow{4}{*}{$f_6$} & Truth & $f_6 = 1.35x_0x_1 + 5.5\sin((x_0-1)(x_1-1))$ \\\cline{2-3}
& \multirow{6}{*}{GP} & $y=((((((x_1\sin(x_0) + x_1x_0 $ \\
&                   & $- \sin(\frac{-x_0}{0.36})-\sin((x_0 + x_1))) $\\
&                   & $- \sin((x_0x_1)^2)) + \sin(\frac{x_1}{0.36}))$\\
&                   & $- \sin((x_1\sin(x_0))^2)) - \sin(\frac{-x_0}{0.36}))$\\
&                   & $- \sin(x_1\sin(x_0)+ x_1x_0)) $\\
&                   & $- \sin(x_1\sin(x_0)+x_1x_0)$ \\ \cline{2-3}
& \multirow{3}{*}{BSR} &$y=(-0.19) + (-0.85)[1.69x_0x_1+1.19]$\\
&                       & $+(7.00*10^{-3})[exp(sin(exp(exp(exp(x_1)$\\
&                       & $+(1.37x_1-1.01)^3)^3)))]$ \\\hline

\end{tabular}
\label{table:expressions}
\end{table}

\subsection{Expressions for Financial data}
Here we present results of BSR on financial data omitted in the paper. Results include training accuracy, testing accuracy and the corresponding expression BSR finds.

\begin{table}[H]\footnotesize
\centering
\caption{Accuracy and Expressions in Real Data Analysis}
\begin{tabular}{|c|c|c|c|}

\hline
\multirow{2}{*}{\bf{\#}}& \multicolumn{2}{|c|}{\textbf{Accuracy}} & \multirow{2}{*}{\textbf{Expression}}\\ \cline{2-3} 
   &    Train & Test &   \\ \hline

\multirow{2}{*}{1} &\multirow{2}{*}{0.539} & \multirow{2}{*}{0.518}&$2.9*10^{-4} - 1.2*10^{-3}*\frac{1}{open^2}$\\
&&& $+1.9*10^{-3}\frac{1}{low^2}$\\ \hline

\multirow{2}{*}{2} &\multirow{2}{*}{0.530} & \multirow{2}{*}{0.525}&$-4.0*10^{-2}- 1.1*10^{-2} *\frac{1}{-2.9*e^{low}+0.84+open}$\\
&&& $-4.0*10^{-2}*\frac{high}{close}$\\

\hline

\multirow{2}{*}{3} &\multirow{2}{*}{0.531} & \multirow{2}{*}{0.501}&$1.3*10^{-3} - 6.4*10^{-5}*e^{close}$\\
&&& $+4.2*10^{-8}*(e^{high})^2$\\

\hline

\multirow{2}{*}{4} &\multirow{2}{*}{0.539} & \multirow{2}{*}{0.518}&$4.2*10^{-4}+ 4.1*10^{-3}*\frac{1}{e^{high^2}}$\\
&&& $-6.1*10^{-5}*e^{\frac{1}{open^2*high}}$\\

\hline

\multirow{2}{*}{5} &\multirow{2}{*}{0.532} & \multirow{2}{*}{0.511}&$1.9*10^{-3} - 5.5*10^{-4}*close$\\
&&& $- 7.2*10^{-7}*(e^{low})^2$\\

\hline

\multirow{2}{*}{6} &\multirow{2}{*}{0.532} & \multirow{2}{*}{0.518}&$5.7*10^{-4}+ 1.6*10^{-4}*\frac{1}{1.1*(open+2*low)+0.43}$\\
&&& $-1.9*10^{-7}*(high^3*e^{low}-high)$\\

\hline

\end{tabular}
\label{table:accuracy_and_expression}
\end{table}

\end{document}